\documentclass{aastex631}

\usepackage{amsmath}

\newcommand{\lyaf}{Ly$\alpha$ forest}
\newcommand{\lya}{Ly$\alpha$}
\newcommand{\hmpc}{\ensuremath{~h^{-1}{\rm Mpc}}}
\newcommand{\hkpc}{\ensuremath{~h^{-1}{\rm kpc}}}

\shorttitle{Constraining FGPA with \lyaf{} tomography}
\shortauthors{Kooistra et al.}
\graphicspath{{./}{figures/}}

\begin{document}

\title{Constraining the Fluctuating Gunn-Peterson Approximation \\ Using Lyman-$\alpha$ Forest Tomography at $z=2$}

\correspondingauthor{Robin Kooistra}
\email{robgonorl@gmail.com}

\author[0000-0002-1008-6675]{Robin Kooistra}
\affiliation{Kavli IPMU (WPI), UTIAS, The University of Tokyo, Kashiwa, Chiba 277-8583, Japan}

\author[0000-0001-9299-5719]{Khee-Gan Lee}
\affiliation{Kavli IPMU (WPI), UTIAS, The University of Tokyo, Kashiwa, Chiba 277-8583, Japan}

\author{Benjamin Horowitz}
\affiliation{Princeton Department of Astrophysical Sciences, Princeton, NJ 08544, USA}
\affiliation{Lawrence Berkeley National Lab, 1 Cyclotron Road, Berkeley, CA 94720, USA}



\begin{abstract}

The fluctuating Gunn-Peterson approximation (FGPA) is a commonly-used 
method to generate mock Lyman-$\alpha$ (Ly$\alpha$) forest absorption skewers at Cosmic Noon ($z\gtrsim 2$) from the matter-density field of
$N$-body simulations without running expensive hydrodynamical simulations.
Motivated by recent developments in 3D IGM tomography observations as well as matter density field reconstruction techniques applied to galaxy redshift samples at $z\sim 2$, we examine the possibility of observationally testing FGPA by directly examining the relationship between the Ly$\alpha$ transmission and the underlying matter density field. Specifically, we analyze the EAGLE, Illustris, IllustrisTNG and Nyx cosmological hydrodynamic simulations, that were run with different codes and sub-grid models.
While the FGPA is an excellent description of the IGM in lower-density regions, the slope of the transmission-density distribution at higher densities is significantly affected by feedback processes causing the FGPA to break down in that regime. Even without added feedback, we find significant deviations 
caused by hydrodynamical effects arising from non-linear structure growth. We then proceed to make comparisons using realistic mock data assuming the sightline sampling and spectral properties of the recent CLAMATO survey, 
and find that it would be challenging to discern between the FGPA and hydrodynamical models with current data sets. However, the improved sightline
sampling from future extremely large telescopes or large volumes from multiplexed spectroscopic surveys such as Subaru PFS should allow for stringent tests of the FGPA, 
and make it possible to detect the effect of galaxy feedback on the IGM.

\end{abstract}



\section{Introduction} \label{sec:intro}
The thermal and ionization states of the intergalactic medium (IGM) are strongly affected by the ionizing radiation fields from galaxies and active galactic nuclei (AGN). 
During the time period between cosmic reionization ($z\sim 7$) and the rise of the warm-hot intergalactic medium (WHIM) starting from $z\sim 1.5$, the vast majority of the IGM resides in photoionization equilibrium with a quasi-uniform background ultraviolet radiation field \citep{cen1994}.  
The optically-thin neutral hydrogen (HI) content of the IGM during this so-called ``Cosmic Noon'' epoch can be probed through the redshifted absorption features seen in the spectra of background sources, especially as the Lyman-$\alpha$ (Ly$\alpha$) forest \citep[e.g.][]{lynds1971,meiksin2009,mcquinn2016}, although higher-order transitions are also detectable \citep{dijkstra2004,irsic2014}. The paradigm of the optically-thin photoionized IGM implies a nearly power-law relationship between the optical depth of the Lyman-$\alpha$ forest and the underlying fluctuations of the matter density field \citep[e.g.][]{cen1994,zhang1995,miraldaescude1996,hernquist1996},
providing a probe of cosmology and large-scale structure at redshifts at which galaxies remain challenging
to observe.\\

The matter density field dependence of the \lyaf{} allows for the possibility of using the Ly$\alpha$ forest at $z\sim 2-4$, or combinations of it with other probes of the density field, to directly constrain cosmological models \citep[e.g.][]{mandelbaum2003, mcdonald2005,seljak2006,palanque-delabrouille2015a,palanque-delabrouille2015b,sainteagathe2019}. However, the evolution of the IGM is strongly affected by heating and feedback processes from galaxies and AGN \citep[e.g.][]{mcdonald2000,mcdonald2005,kollmeier2006,arinyoprats2015,tonnesen2017}, which implies feedback would affect the connection between the \lyaf{} and its underlying density field. Fortunately, the effect of different models of galaxy and AGN feedback on the global statistics of the Ly$\alpha$ forest, such as the probability distribution function (PDF) and the 1-dimensional transmission power spectrum, appear to be only of the order of a few percent (e.g., \citealt{lee:2011}, \citealt{viel:2013b}, \citealt{chabanier:2020}). Nevertheless, the effect can be expected to be stronger in regions with large-scale overdensities of a few times the cosmic mean density that lie in the vicinity of the galaxies and AGN from which the feedback originates \citep[e.g.][]{rakic2013,meiksin2015,meiksin2017,turner2017,sorini2018,sorini2020,kooistra2019,preheatingpaper}.\\

An example of this was shown in \citet{nagamine2021}, who examined the effect of several models of star formation and supernova feedback on the one-dimensional (1D) statistics of the Lyman-$\alpha$ forest. They specifically looked at the observables mentioned above in the context of existing surveys, as well as the upcoming Subaru Prime Focus Spectrograph (PFS) survey \citep{pfs}, and found that the different feedback prescriptions particularly have an effect on the \lya{} transmission contrast in the vicinity ($\lesssim$Mpc) of galaxies. More recently, \citep{preheatingpaper} carried out a study of simulated 3D \lyaf{} tomographic maps covering galaxy proto-clusters, i.e.\ mildly overdense progenitor regions of modern-day galaxy clusters. We found that the distribution of the \lya{} transmission versus the dark matter (DM) density (first considered by \citealt{hyperion1}) is highly sensitive to different levels of energy injection induced by
non-gravitational heating of the proto intra-cluster medium. This suggests that the different models of feedback adopted in cosmological simulations could also affect this distribution. Such a study, however, would require a measurement of the underlying DM density field within a survey volume coinciding with the 
Lyman-$\alpha$ absorption.\\

This has recently become feasible with two recent innovations in the study of the IGM and large-scale structure at the $z\sim 2-3$ era of Cosmic Noon. The first is the development of 3D Lyman-$\alpha$ forest tomography. The Lyman-$\alpha$ forest has traditionally been observed in the foreground of quasars, typically at high spectral resolution ($R \equiv \lambda/\Delta\lambda \gtrsim 10^4$) and high signal-to-noise (S/N$\gtrsim$10 per resolution element). Such data can only be obtained with bright quasars which are relatively rare, such that the individual sight lines can be studied only as one-dimensional probes of the IGM. However, it has been shown that by targeting much fainter star-forming galaxies in addition to quasars, it is possible with existing observational facilities to observe a sample of \lyaf{} sightlines with high areal density within specific regions on the sky \citep{lee2014a,lee2014b}. While the individual spectra are of low signal-to-noise (S/N $\sim$ few per resolution element) and moderate spectral resolution ($R\sim 10^3$), the information in these skewers can be combined to construct accurate three-dimensional (3D) maps of the large-scale HI absorption field through Wiener-filtering \citep{pichon2001,caucci2008}. This technique has become known as \lyaf{} or IGM tomography and has been pioneered by the COSMOS Lyman Alpha Mapping And Tomographic Observations (CLAMATO) survey \citep{clamato,clamato2}. This large-scale tomographic survey probes the well-studied COSMOS field with 360 galaxy and quasar sightlines over an area of $\sim0.2$ deg$^2$ and a  redshift range of 2.05 $<$ $z$ $<$ 2.55. Various works have demonstrated the utility of using such a survey to study large-scale structure and the formation of proto-clusters at Cosmic Noon \citep[e.g.,][]{hyperion1,hyperion2,krolewski2018}. More recent efforts include that of \citet{ravoux2020}, as well as the Lyman Alpha Tomography IMACS Survey \citep[LATIS;][]{latis}, both covering larger areas of 220 deg$^2$ and 1.7 deg$^2$, respectively, but with larger sight line separations than CLAMATO. The upcoming Subaru Prime Focus Spectrograph will also carry out \lyaf{} tomography over 12 deg$^2$ as part of its Subaru Strategic Survey program 
(see the Appendix of \citealt{nagamine2021}). \\

The second, more recent, development is the application of matter density field reconstruction techniques on high-redshift galaxy spectroscopic survey data \citep[e.g.,][]{tardis1,birthcosmos,tardis2}. In \citet{birthcosmos}, high-redshift data from multiple spectroscopic galaxy surveys were combined, along with a galaxy bias prescription, to reconstruct a 3D matter density field of the COSMOS field using constrained realizations. This resulted in a 3D map of the matter density within the portion of the COSMOS field, with an effective spatial resolution of $\sim$5 Mpc over a redshift range of $1.6<z<3.2$ (see also Ata et al, in prep.). This dataset can therefore uniquely be combined with the tomographic \lyaf{} data from CLAMATO, which covers the same cosmic volume, in order to study the effects of heating on the intergalactic gas \citep[see also][]{preheatingpaper} specifically in proto-cluster overdensities.\\

In this paper, we generalize the idea of directly studying the link between \lya{} transmission and underlying matter density first laid out in \citet{preheatingpaper}. Instead of studying just the regions surrounding galaxy protocluster overdensities as was done in the earlier paper, we  propose to use the observed \lya{} transmission - DM density distribution across a wide density range as a direct test of the commonly-used fluctuating Gunn-Peterson approximation \citep[FGPA; e.g.,][]{hui1997a,croft:1998,weinberg:2003,rorai:2013}. The FGPA\footnote{Not to be confused with FPGA, or field-programmable gate array, which is a type of integrated circuit.} is an analytic approximation that is based on an equilibrium between optically-thin photoionization and collisional recombination of residual HI in the IGM, and leads to a power-law relation between the density field and the resulting optical depth:

\begin{equation}
\tau \propto \left(1+\delta\right)^{\beta}.\label{eq:FGPA}
\end{equation}

Here $\delta$ denotes the matter overdensity and is defined as $\delta \equiv \rho/\bar{\rho} - 1$. The power-law slope $\beta$ is often related to the slope of the temperature-density relationship $\gamma$ through $\beta = 2-0.7\left(\gamma - 1\right)$, where $\gamma$ follows from the "equation of 
state" of the IGM:

\begin{equation}
T \propto \left(\rho/\bar{\rho}\right)^{\gamma-1},
\end{equation}

which depends on the photoionization equilibrium of the gas, as well as the heating mechanisms affecting it \citep{hui1997b,croft:1998,weinberg:2003}. Although this relation is based on the baryon density field, it can also be applied to the DM density field to generate mock \lya{} optical depth skewers that match well the properties of skewers calculated using full hydrodynamics \citep[e.g.,][]{sorini2016}. A generally chosen value for the index of the IGM temperature-density relationship, based on observations of Lyman-$\alpha$ forest statistics, is $\gamma = 1.6$ \citep[e.g.,][]{rudie2012,bolton2014,lee2015,hiss2018}. This yields an FGPA slope (between the \lya{} transmission overdensity and matter overdensity) which is coincidentally also $\beta\approx 1.6$; this fiducial value will be adopted throughout this work. The FGPA is generally able to reproduce \lyaf{} statistics, such as the 1D line power spectrum and transmission PDF, that are consistent with those seen from hydrodynamical simulations on large scales and in mean- to low-density regimes making up the majority of the cosmic volume of the IGM \citep[see e.g., Figure 12 in][]{sorini2016}. However, the assumption of a power-law temperature-density relation tends to break down for strongly heated gas as well as at higher densities \citep[e.g.,][]{lukic2015}. A recent comparison between \lyaf{} tomographic maps based on the FGPA and full hydrodynamical maps was made by \citet{qezlou2021}, focusing on the characterization of proto-cluster masses. They found overall good agreement between the two, although some of the detailed structures in the tomographic maps were different.\\

In this work, we explore how well the FGPA holds up in comparison with several publicly-available hydrodynamical simulations in the context of the potentially observable \lya{} transmission-DM density distribution. We generate mock \lya{} skewers from multiple cosmological hydrodynamical simulations, each adopting a different feedback model. We then compare the behavior of the transmission-density distribution for the DM-based FGPA with the transmission-density distributions derived in the presence of baryonic physics and heating due to stars and AGN, based on the hydrodynamical simulations. In particular, we focus on the slope of the \lya{} transmission-DM density distribution at higher DM densities.\\

An introduction to each of the simulations is given in Section \ref{sec:sims}. This is followed by a detailed description of the methodology adopted to generate the \lya{} tomographic maps required for the \lya{}-DM density distribution in Section \ref{sec:tomogr}. The resulting distributions are then presented and discussed in Section \ref{sec:results}.

\section{Simulations}\label{sec:sims}
In order to be able to study the IGM in detail, a hydrodynamical simulation with a large cosmological volume is preferential.
However, this needs to be traded-off with the need for a grid resolution that is small enough to model the detailed physics occurring on the sub-grid scales governing processes such as star and black hole formation, star formation feedback, AGN feedback etc. Since all these processes affect the temperature and ionization state of the IGM in a different manner, different models should result in distinct \lya{} transmission distributions as a function of the underlying matter density. For the lower-density IGM gas that resides far from most galaxies and AGN, the effect of these feedback processes should be minimal and most of the gas will follow the power-law temperature-density relationship set by photoionization equilibrium with the permeating ultraviolet background, resulting in the standard FGPA distribution in transmitted flux. However, in the higher-density IGM gas, closer to galaxies, the specific prescriptions for feedback and baryonic physics in the simulation will have a significant impact on the HI density and thus the \lya{} transmission. Therefore, we use a set of multiple simulations run with different codes and models in order to study how they affect the \lya{} transmission-DM density distribution at $z$ = 2. A brief description of these simulations follows in the sections below and a summary of their properties can be found in Table \ref{tab:sims}. For every simulation we adopt the respective cosmology that was used to run it in our calculations. The cosmological parameters relevant for this work can also be found in Table \ref{tab:sims}.
As an aside, we note that while these simulation boxes are all commonly accepted as 'cosmological simulations', their volumes are still too small for most actual cosmological analyses. Most pertinently, the simulation volumes are insufficient for encompassing at least several massive Coma-like galaxy clusters with $M \sim 10^{15}\,\mathrm{M_{\odot}}$, which is why these simulations were not used for the protocluster study in \citet{preheatingpaper}.
Nevertheless, the volumes of the four simulations in Table \ref{tab:sims} are adequate for the study of less extreme density regimes that make up the vast bulk of cosmic volume.

\begin{table}
  \centering
	\caption{Hydrodynamical simulation properties. The length of a side of each simulation box is denoted by $L_{\rm box}$. The cosmological parameters listed here are the Hubble parameter $h$, the matter energy density $\Omega_{\rm m}$ and the baryon energy density $\Omega_{\rm b}$.}
	\label{tab:sims}
    \hspace*{-2cm}
	\begin{tabular}{lccccc}
		\hline
		Simulation suite & Simulation model & $L_{\rm box}$ ($h^{-1}$Mpc) & Cosmology ($h$, $\Omega_{\rm m}$, $\Omega_{\rm b}$) & Star formation & Feedback types\\
		\hline
		Nyx & L100\_N4096 & 100 & (0.685, 0.3, 0.047) & No & None \\ 
		& & & & & \\
		Illustris & Illustris-1 & 75 & (0.704, 0.2726, 0.0456) & Yes & SN galactic winds; \\
		& & & & & Radiative \& mechanical AGN\\ 
		IllustrisTNG & TNG100-1 & 75 & (0.6774, 0.3089, 0.0486) & Yes & SN galactic winds;\\
		& & & & & Radiative \& mechanical AGN \\
		EAGLE & RefL0100N1504 & 67.77 & (0.6777, 0.307, 0.04825) & Yes & Thermal stellar \& \\
		& & & & & Thermal AGN \\ 
		\hline
	\end{tabular}
\end{table}

\subsection{Nyx}
The largest volume simulation we adopt was created using the Eulerian hydrodynamical code Nyx \citep{nyxsim,lukic2015}. This simulation box has a volume of $\left(100 h^{-1}\mathrm{Mpc}\right)^3$. Although Nyx allows for adaptive mesh refinement, this particular simulation was run without it since we are most interested in the diffuse IGM component. The hydrodynamical properties were calculated using a fixed grid with 4096$^3$ cells and as many DM particles \citep{enigmadescr,sorini2018}. The high resolution allows for a detailed study of the IGM across the entire volume and recovers sufficiently small scales for the \lyaf{} with the corresponding 24 \hkpc{} resolution \citep{lukic2015}. Neither star formation nor galaxy feedback models were included in the Nyx simulation. This Nyx simulation can thus be considered a reference hydrodynamical model where only general baryonic physics, such as baryonic pressure, affect the evolution of the optical depth, independent of heating or feedback mechanisms that can influence the properties of the IGM closer to galaxies. The specific physics of the cooling, recombination, collisional ionization, and dielectric recombination rates are described in \citep{lukic2015}, with the ionizing ultraviolet background described in \citet{2012Haardt} and self-shielding described in \citet{rahmati2013}. The simulation is initialized at $z_{\rm ini} = 200$ and evolved until the $z$ = 2 snapshot.

\subsection{Illustris}
The Illustris suite of simulations are based on the Voronoi moving-mesh code \citep[\textsc{Arepo}:][]{arepo,arepo_public} where the dark matter particles are treated with a Lagrangian approach while the baryons are evolved as an ideal gas on a moving  Voronoi tessellation grid. The dark matter evolution uses the TreePM \citep{1995xu}, a method that uses a particle-mesh code \citep{1981csup.book.....H,1997astro.ph.12217K} for long-range interactions while short-range interactions are handled by a separate hierarchical algorithm \citep{1986Barnes}. We adopt the largest volume box with the highest resolution dubbed Illustris-1, which has a box length of $L$ = 75 \hmpc{} on a side \citep{illustrisa,illustrisb,illustrispub}. We focus on snapshot 68, corresponding to $z$ = 2. This volume is filled with 1820$^3$ DM particles and an equal number of gas elements, resulting in initial masses of $m_{\rm DM}$ = $6.3\times10^6 M_\odot$ for the DM particles and $m_{\rm g}$ = $1.3\times10^6 M_\odot$ for the gas elements.\\

The Illustris simulations do include star and black hole formation and should therefore be more realistic than Nyx. For the star formation model, the \citet{2003Chabrier} initial mass function was used and starforming gas was modeled using an effective equation-of-state with a fixed hydrogen number density threshold for star formation of 0.13 cm$^{-3}$. Additionally, the Illustris model includes descriptions for kinetic feedback by galactic winds due to supernovae. This is implemented both globally throughout the star forming gas where particles are launched freely in a random direction until they reach a density threshold, as well as locally by stochastically launching gas particles that receive energy from nearby stellar particles. For AGN feedback, Illustris employs a two-state model. In the quasar-mode, energy is released radiatively by heating the gas close to the central black hole within a fixed mass scale. The mechanical radio-mode model is based on \citet{sijacki2007}, with particular parameters set by matching smaller scale simulations with observed data \citep{2011Gu,2013Behroozi}.\\

The feedback model adopted in Illustris is known to produce too much heating of the IGM gas, which results in very high temperature gas around galaxies and strongly reduces the small-scale structure \citep[e.g.,][]{genel2014}. A comparison between Nyx and Illustris was made in \citet{sorini2018}, showing that the additional feedback included in Illustris only mildly affects the global temperature-density relationship. However, in the vicinity of quasars and galaxies they found the temperature-density relationship is affected more strongly by the adopted feedback model and consequently results in variations in the \lya{} transmission contrast.

\subsection{IllustrisTNG}
IllustrisTNG is a suite of simulations based on Illustris, which adopts an updated subgrid model to relieve tensions with multiple observations and implements improvements in the hydrodynamical methods \citep{tng1,tng2,tng3,tng4,tng5,tngpub}. These improvements include the inclusion of magnetic fields, an updated spatial gradient algorithm, improved advection of metal abundances, as well as specific galaxy physics described below. The TNG100-1 simulation we adopt here uses the same initial conditions as Illustris-1, but applies updated cosmology (see Table \ref{tab:sims}), as well as the updated model prescriptions \citep{tngmodel1,tngmodel2}. For this work, we analyzed snapshot 33, again corresponding to $z$ = 2. The $L$ = 75 \hmpc{} box contains 1820$^3$ DM and gas particles, where the initial mass of the latter is $m_{\rm g}$ = $1.4\times10^6 M_\odot$ and the DM particles have a mass resolution of $m_{\rm DM}$ = $7.5\times10^6 M_\odot$.\\

The galactic wind model in TNG100-1 is similar to that used in Illustris-1, but now the particles are ejected isotropically and the ejection mechanism has been updated with more recent advancements \citep[see][]{tngmodel2}. Similarly, the AGN feedback model in IllustrisTNG also contains largely the same ingredients as in Illustris, but employs updated prescriptions for the kinetic and thermal energy injection that help resolve some of the excessive heating found with the Illustris model \citep[see][]{tngmodel1}.\\

Since TNG100-1 implements the same initial conditions as Illustris-1 it allows for direct comparison between structures within the simulations, where the only differences affecting the IGM will be due to the cosmologies and the different feedback models.

\subsection{EAGLE}
The final simulation volume we consider is part of the Evolution and Assembly of GaLaxies and their Environments (EAGLE) suite of simulations \citep{eagle1,eagle2}. EAGLE is based on a modified version of the Smoothed Particle Hydrodynamics (SPH) code Gadget-3 \citep{gadget}.  We selected the $z$ = 2 snapshot (snapshot 15) from the largest box available in the public data release, RefL0100N1504, which has a box length of $L_{\rm box}$ = 67.77 \hmpc{} \citep{eagle_pub}. It was run using 1504$^3$ DM particles with the same number of baryonic particles, resulting in a DM particle resolution of $m_{\rm DM}$ = $9.70\times10^6 M_\odot$ and an initial gas particle mass of $m_{\rm g}$ = $1.81\times10^6 M_\odot$.\\ 

Like the previous two simulations, EAGLE includes subgrid physics prescriptions for the formation of stars and black holes \citep{eagle1}. Star formation is triggered in particles with the metalicity-dependent density threshold from \citet{schaye2004} following the  Kennicutt-Schmidt law \citep{kennicutt1998} with a Chabrier initial mass function \citep{chabrier2003}. Stellar feedback from supernovae is implemented stochastically by increasing the temperature by $\Delta T_{*}$ =  10 K when stellar particles reach an age of $3 \times 10^7$ yr \citep{dallavecchia2012}. AGN feedback was implemented by setting a heating temperature $\Delta T_{\rm AGN}$ = 10$^{8.5}$ K of the central black hole particle and determining a probability for heating the neighboring simulation particles by calibrating the feedback efficiency on observations of the galaxy size – stellar mass relation, the black hole mass – stellar mass relation and the galaxy stellar mass function \citep[see][]{eagle2,schaye2015}.

\section{\texorpdfstring{Lyman$\alpha$ transmission-dark matter density distributions}{Lyman alpha transmission-dark matter density distributions}}\label{sec:tomogr}
For each of the simulations, we aim to study the distribution of the \lya{} transmission as a function of the DM density field. In order to make a fair comparison, we treat all the simulations in the same manner. This section describes how we generate the mock \lya{} skewers that we use to create the tomographic maps. The output data from every simulation comes in a different format. The Illustris and IllustrisTNG simulations, as well as EAGLE come with the full particle data. For the former two, the particles represent the Voronoi cells, whereas in EAGLE they represent a cloud of SPH particles. All fields in the output of Nyx, however, already come rendered onto a regular grid with $N$ = 4096$^3$ cells or a resolution of $\sim$24 comoving \hkpc. Therefore, before calculating the \lya{} optical depth in the other simulations, we also gridded their baryon fields (density, temperature and velocity along the z-axis) to a regular grid. For Illustris-1 and TNG100-1, we work with $N$ = 600$^3$ cells, corresponding to a comoving resolution of 125 \hkpc. Likewise, we deposited the EAGLE particles onto a grid with the same $\sim$125 \hkpc{} resolution by adopting $N$ = 542$^3$ grid cells. The gridding itself was performed using the \verb|yt| package \citep{ytpython} in gather smoothing mode. We note that ideally the gridded resolutions of the other simulations should be the same as for Nyx. However, the gridding operation proved to be highly memory intensive and we were unable to achieve that resolution with our available computing resources. Additionally, rather than throwing away data by re-binning the grid to $N$ = 800$^3$ cells, which would also introduce errors, we settled on keeping the full 4096 cells per skewer for Nyx. We also note that an increased resolution for the other simulations at this stage would not affect our conclusions due to smoothing at a later stage.\\

In EAGLE, the particle properties are defined following an SPH kernel assuming 58 nearest neighbors \citep{eagle1}. In the case of Illustris(TNG), the baryon fields were defined within each individual voronoi cell, hence an SPH kernel with many nearest neighbors would poorly represent the voronoi kernel. Therefore, we adopted  a cubic SPH kernel for the gridding with the original 58 nearest neighbors for EAGLE, but used only 4 nearest neighbors for Illustris(TNG). The number for Illustris(TNG) was found to yield the most accurate results with regards to mass conservation of the density field. The smoothing length $h_{\rm sm}$ was then calculated from the particle density and mass, assuming spherical geometry as:

\begin{equation}
h_{\rm sm} = \left(\frac{3}{4\pi}\frac{m_{\rm part}}{\rho_{\rm part}}\right)^{1/3},
\end{equation}
where $m_{\rm part}$ denotes the particle mass and $\rho_{\rm part}$ its density. 

\subsection{Hydrodynamical skewers}\label{sec:hydroskew}
From the gridded baryon fields we then extracted skewers along the z-axis. Due to the high resolution of the Nyx simulation grid, we extracted 800$\times$800 equally spaced skewers along the x and y axes, keeping the full 4096 pixels in the z-direction. We then use the same methodology as in \citet{enigmadescr} and \citet{lukic2015} to determine the \lya{} optical depth. In short, the HI density along each skewer was calculated from the temperature and density by assuming photoionization equilibrium. The equilibrium rate equations were solved with the \citet{hm12} photoionizing background providing the photoionization rate and including the effect of self-shielding \citep{rahmati2013} to obtain the HI number density $n_{\rm HI}$. We note that using the same UV background model for each simulation is not entirely self-consistent. However, the effect of this is later negated by renormalizing the transmission. The influence of the feedback models in each simulation on the HI number density should then come into play solely through the gas temperature, taken directly from the gridded simulations.\\

Subsequently, the \lya{} optical depth in redshift space was determined as follows:

\begin{equation}
    \tau_{\rm Ly\alpha} = \int n_{\rm HI}\,\sigma_\nu \mathrm{d}r.
\end{equation}

Here, $\mathrm{d}r$ denotes the line element along the line of sight and $\sigma_{\rm v}$ denotes the cross-section, which includes thermal Doppler broadening with a Voigt line profile; the full details can be found in \citet{lukic2015}. 
From there, the \lya{} transmission $F$ is defined as

\begin{equation}
    F = e^{-\tau}.
\end{equation}

Before further processing, the mock skewers in every simulation were normalized to reproduce the observed mean \lya{} transmission from \cite{meanflux} and we adopt the usual convention of the transmission overdensity defined by $\delta_{\rm F} \equiv F/\bar{F} - 1$, where $\bar{F}$ denotes the mean transmission. The final skewers were put back onto a cubic grid to generate the 3D transmission maps. Examples of single skewers for all the simulations are shown in Section \ref{app:singskew} in the appendix. Moreover, commonly used \lyaf{} statistics, such as the PDF of the transmission and the 1D line powerspectra can be found in Sections \ref{app:fluxpdf} and \ref{app:pspec}, respectively.
 
\subsection{Redshift-space DM density fields}\label{sec:rsddm}
The DM density is required in order to be able to calculate the FGPA optical depth. For Illustris, IllustrisTNG and EAGLE we deposited the DM particles onto a regular grid with the same number of cells as the baryon fields described above, using cloud-in-cell (CIC) interpolation. For Nyx, the DM density field was resampled along the same 800$\times$800 sight lines as for the hydrodynamical skewers.\\

For the final \lya{} transmission-DM density distribution, we furthermore convolved the density fields with peculiar velocities to convert them to redshift space for consistency with the redshift-space optical depth. This is representative of observations, where both the optical depth and DM density field reconstructions would be obtained in redshift space. In the case of Illustris, IllustrisTNG and EAGLE, the particles were shifted along the z-axis by $\Delta v = v_{\rm z}(1+z)/H(z)$ before applying the 3D CIC gridding. The particle positions were then wrapped around the simulation edges to take into account the periodic boundary conditions. Similarly for Nyx, the cells along each skewer were treated as particles, which were then shifted in position based on their corresponding line-of-sight velocity. These were then deposited back onto the regular grid using 1D CIC interpolation.

\subsection{FGPA skewers}\label{sec:fgpaskew}
In addition to the hydrodynamical skewers, we also generated mock skewers from each DM density field using the FGPA. The real space FGPA optical depth was determined through Equation \ref{eq:FGPA} and consequently these skewers were convolved with the velocity field along the skewer axis, where thermal broadening was included through the following integral

\begin{equation}
    \tau_{\rm FGPA} = A_{\rm norm}\cdot \int \left(1+\delta_{\rm DM}\right)^\beta \,\frac{1}{b}\, H_{\rm V}\,(\frac{v_{0}-v}{b})\cdot \mathrm{d}v,
\end{equation}

with $b = \sqrt{2k_{\rm B}T/m_{\rm p}}$ the Doppler parameter, $v_0$ the real-space velocity of the cell and $A_{\rm norm}$ the normalization constant to reproduce the observed mean flux. This includes the full Voigt line-profile convolution and the Voigt function $H_{\rm V}$ is defined as 

\begin{equation}
    H_{\rm V}(x) = 
    \begin{cases}
    1-\frac{2}{\sqrt{\pi}}\frac{\lambda_{\rm Ly\alpha}A_{10}}{4\pi} & \left|x\right|\leq10^{-4}\\
    e^{-x^2} - \frac{\lambda_{\rm Ly\alpha}A_{10}}{4\pi}\frac{1}{\sqrt{\pi}x^2}\cdot\left(e^{-2x^2}\left[4x^4 + 7x^2 + 6x^{-2}\right]-\frac{3}{2}x^{-2} - 1\right) & \left|x\right|>10^{-4}
    \end{cases},
\end{equation}

where the parameter $x = (v_0-v)/b$, $\lambda_{\mathrm{Ly\alpha}}$ = 1215.67 \text{\AA} the Ly$\alpha$ rest wavelength, and $A_{10}$ is the Einstein A coefficient \citep{voigt}. 
Before performing the integral, we first smooth both the DM velocity field and the DM density field to a comoving scale of 228 kpc in order to mimic the baryonic pressure smoothing in the IGM, as was explored in \citet{sorini2016} (but see also \citealt{kulkarni:2015}).\\

\begin{figure}
    \centering
	\includegraphics[width=0.95\columnwidth]{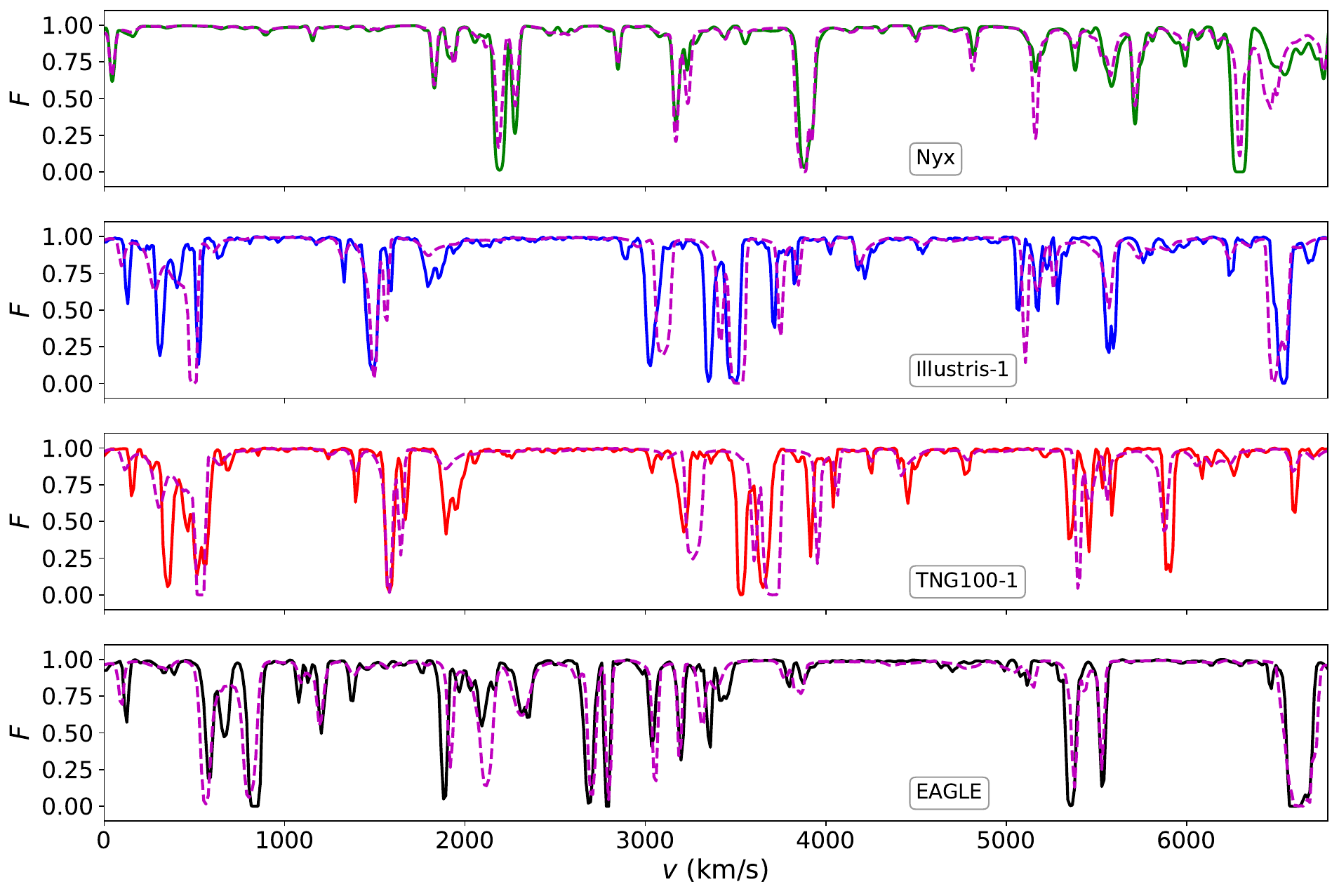}
	\caption{Examples of \lyaf{} skewers through each of the simulations. The solid colored lines denote the hydrodynamical skewers, whereas the magenta dashed lines show the FGPA-based skewers generated along the same sightlines.}
    \label{fig:singskew_all}
\end{figure}

We note that the DM velocities were not saved in the output of the Nyx run we are using and hence we use the baryon velocities directly for Nyx instead. As was noted in \citet{sorini2016}, the optimal smoothing scale would be different for every simulation and hence it is better to use the baryon density and velocities in the FGPA equation instead. However, observationally the DM density is more easily accessible. We therefore opted to apply the constant smoothing to the DM density for EAGLE, Illustris and IllustrisTNG and used the baryon velocities without smoothing for Nyx. In Figure \ref{fig:singskew_all} we present a comparison between examples of hydrodynamical and FGPA skewers for every simulation. It shows that the FGPA overall captures roughly the same features as the hydrodynamical skewers at the same positions, but the details of the lines, such as the depth of the absorption and the width of the lines differ. As with the hydrodynamical skewers, a 3D \lya{} transmission map was constructed for the FGPA by placing the skewers back to their location in the 3D volume along the z-axis.

\subsection{Mock Observational IGM Maps}\label{sec:clamatonoise}
Realistically, the \lyaf{} skewers from the simulations have higher resolution than what is feasibly obtained in observations. Moreover, we thus far have not included any instrumental noise. In order to see if the results will still hold up under these conditions, we additionally create skewers that are smoothed and rebinned to the same spectral resolution, and have had noise introduced at a level consistent with the \lyaf{} data from the CLAMATO survey \citep{clamato,clamato2}. Another possibility we examine is IGM tomography surveys using the future thirty-meter class extremely large telescopes (ELTs), namely the Thirty-Meter Telescope \citep{skidmore:2015}, European Extremely Large Telescope \citep{evans:2014}, and Giant Magellan Telescope \citep{johns:2012}. These powerful facilities will target considerably fainter background sightlines than CLAMATO, with the concomitant increase in sightline density \citep{lee2014a}.\\ 

The noise is added in the same manner as was described in \citet{tardis1} and \citet{tardis2}. For ELT-like noise we adopt the parameters from \citet{tardis1}. In more detail, we performed the following steps:

\begin{enumerate}
    \item Normalize the optical depth to the observed mean flux from \citet{meanflux}, which follows the fitting function $\bar{\tau} = 0.001845\left(1+z\right)^{3.924}$.
    \item Smooth the spectra to the instrumental resolution of 4 \text{\AA} full width at half maximum (FWHM) and rebin them to the pixel size of 1.2 \text{\AA} as sampled by the Keck-I/LRIS spectrograph used by CLAMATO. For ELT, the resolution was set to be three times higher, yielding a resolution with FWHM of 4 \text{\AA} and a pixel size of 0.4 \text{\AA}.
    \item For every skewer, draw a random number from a power-law distribution representing the signal-to-noise (S/N) to be applied to the given skewer. The power-law slope was fixed to 2.7. The minimum S/N was then capped to 0.2 to be representative of the data from the CLAMATO observations \citep{stark:2015a,krolewski2018} and to 2.8 for ELT-like mocks.
    \item Add continuum errors based on the S/N values to each of the skewers. To this end, the transmission of the skewers was modified as $F_{\rm cont} = F_{\rm sim}/\left(1+\delta_{\rm cont}\right)$. Here the $\delta_{\rm cont}$ is a continuum error drawn from a Gaussian distribution with standard deviation $\sigma_{\rm cont} = 0.2054/(S/N)+0.015$ following \citet{krolewski2018}. The continuum error was capped at a S/N value of 10.
    \item Add Gaussian noise to the continuum corrected skewers based on the previously obtained S/N.
    \item Randomly select a number of skewers to obtain a similar mean sightline density to CLAMATO or ELT. In this case, we drew $(L/2.5\, h^{-1} \mathrm{Mpc})^2$ skewers for each simulation for CLAMATO, resulting in 1600, 900, 900 and 734 skewers for Nyx, Illustris, IllustrisTNG and EAGLE, respectively. For ELT this corresponds to $(L/1.0\, h^{-1} \mathrm{Mpc})^2$, or 9999, 5625, 5625 and 4592 skewers, respectively.
\end{enumerate}

Finally, the resulting skewers were processed with the Wiener-filtering code {\verb|dachshund|}\footnote{https://github.com/caseywstark/dachshund} described by \citet{stark:2015a} in order to obtain 3D tomographic maps of the \lyaf{} transmission, that are realistic representations of observations with a survey such as CLAMATO or an ELT. 
Note that in terms of cosmic volume, CLAMATO has sampled a comoving volume of $V = 4.1 \times 10^5\,h^3\;\mathrm{Mpc}^{-3}$ \citep{clamato2}, which is $2.5\times$ smaller than the $V=10^6\,h^3\;\mathrm{Mpc}^{-3}$ volume sampled by the Nyx box but comparable to the EAGLE and Ilustris/IlustrisTNG simulation volumes ($V=3.1 \times 10^5\,h^3\;\mathrm{Mpc}^{-3}$ and $V=4.2 \times 10^5\,h^3\;\mathrm{Mpc}^{-3}$, respectively). We therefore expect any variance from the different CLAMATO-like mock realizations to be dominated by the sightline sampling and pixel noise, rather than from sample variance caused by the finite volume. 

\begin{figure*}[bt!]\centering
	\includegraphics[width=0.49\columnwidth]{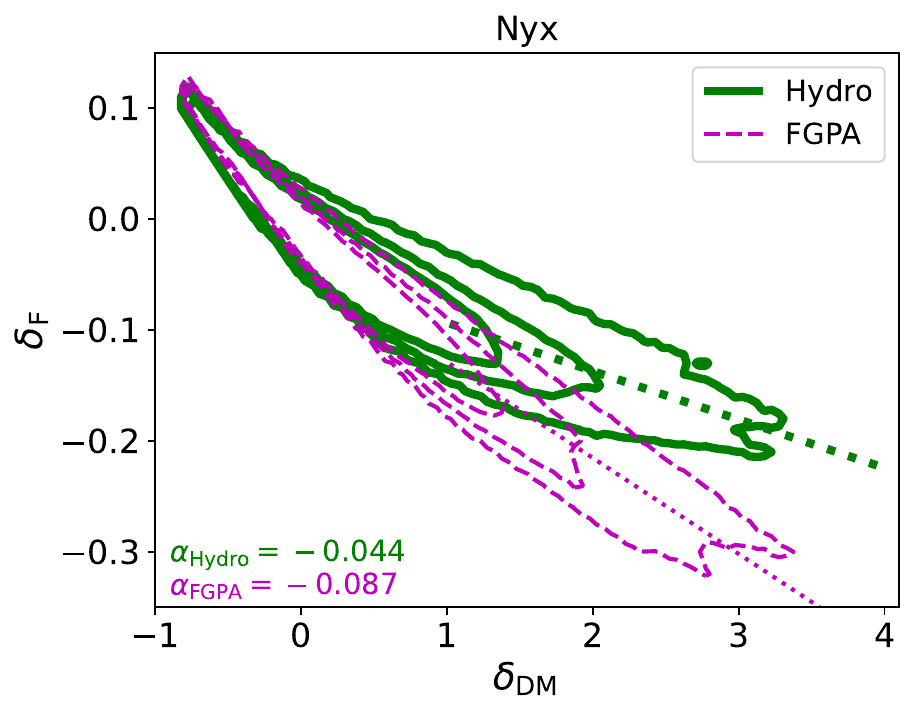} 
	\includegraphics[width=0.49\columnwidth]{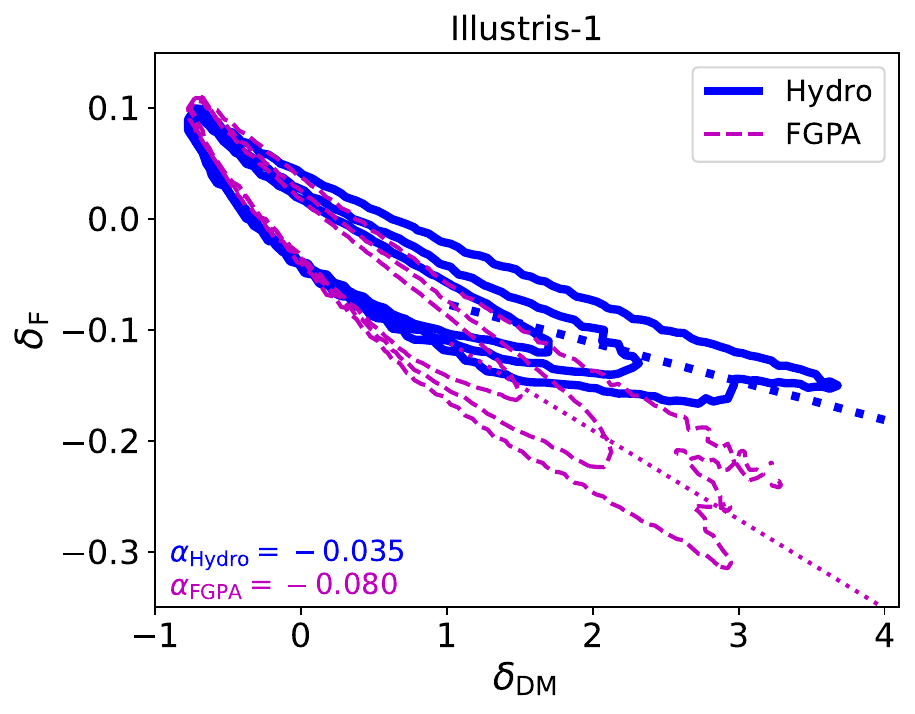}
	\includegraphics[width=0.49\columnwidth]{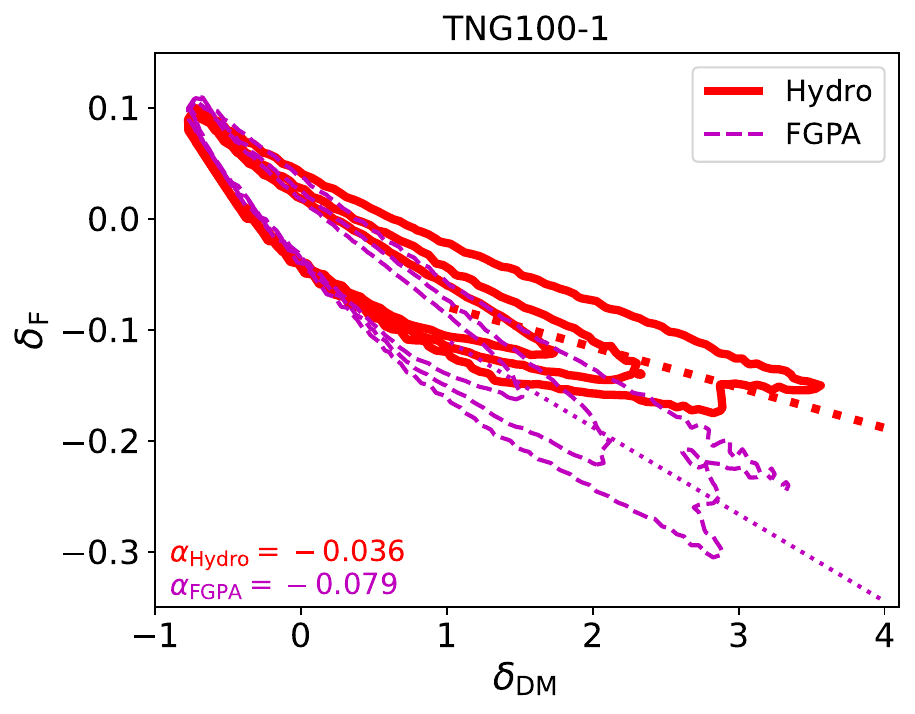} 
	\includegraphics[width=0.49\columnwidth]{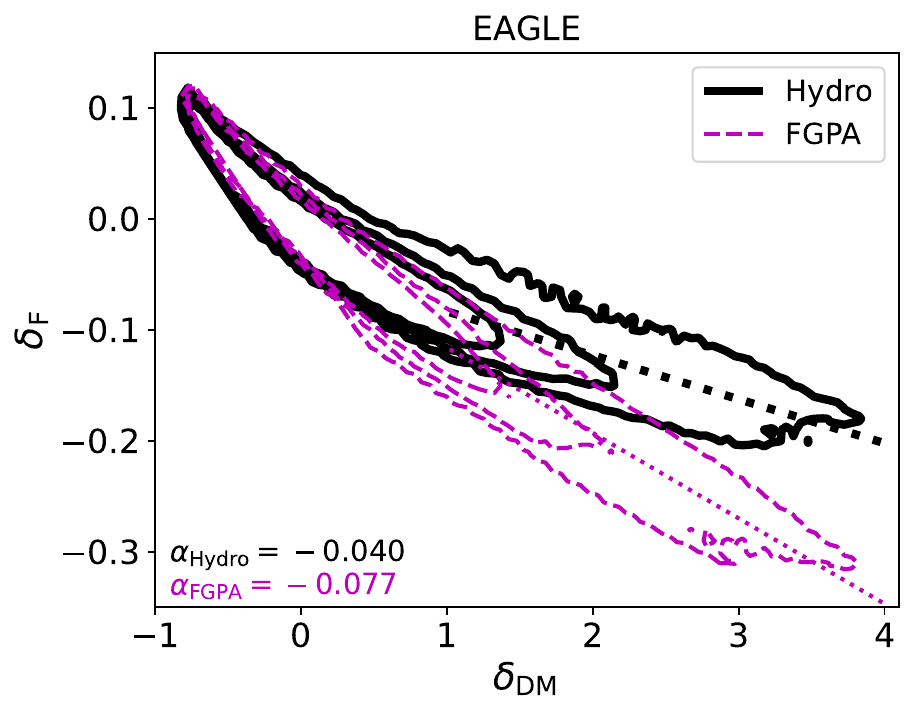}
	\caption{Ly$\alpha$ transmission - DM density distribution of all simulations compared to their respective FGPA-based distributions. The contours denote the 2\%, 20\% and 80\% levels of the PDF of the distribution, respectively. The thick dotted line denotes the fit to the slope of the distribution based on the hydrodynamical skewers and the thin dotted line shows the same for the FGPA skewers. The values of the slopes $\alpha$ are written in the bottom left corner of every panel.}
    \label{fig:lyadm_indiv}
\end{figure*}

\section{Results and discussion}\label{sec:results}
In this section we present the \lya{} transmission-DM density distributions following the methodology described previously in Section \ref{sec:tomogr} and analyze them in detail. 

\subsection{Simulated Transmission-Density Distributions}\label{sec:sim_dist}
In order to visualize the \lya{} transmission-DM density distribution and highlight the differences more clearly, we first smoothed both the raw tomographic \lya{} transmission maps (see Section \ref{sec:hydroskew}) and the redshift-space DM density fields using a Gaussian kernel with standard deviation of $\sigma = 3\, \hmpc$. The kernel size is similar to the maximum effective smoothing scale of the CLAMATO survey data \citep{clamato}. We performed the same for the tomographic maps based on the FGPA skewers (see Section \ref{sec:fgpaskew}).\\ 

The resulting hydrodynamical transmission-density distribution for each of the simulations described in Section \ref{sec:sims} is shown in Figure~\ref{fig:lyadm_indiv}. The contours represent the 2\%, 20\% and 80\% levels of the PDF of the distribution. The FGPA distributions are given by the magenta dashed contours, whereas the solid colored contours indicate the full hydrodynamical distribution. As can be seen in the figure, the distribution generally follows a banana-shaped curve, similar to what was seen in \citet{hyperion1}.  However, the hydrodynamical skewers curve up more towards higher transmission (less absorption) at high matter densities than their FGPA equivalents, whereas the distributions at low densities are roughly consistent with each other. \\

\begin{figure}[bt!]
    \centering
	\includegraphics[width=0.65\columnwidth]{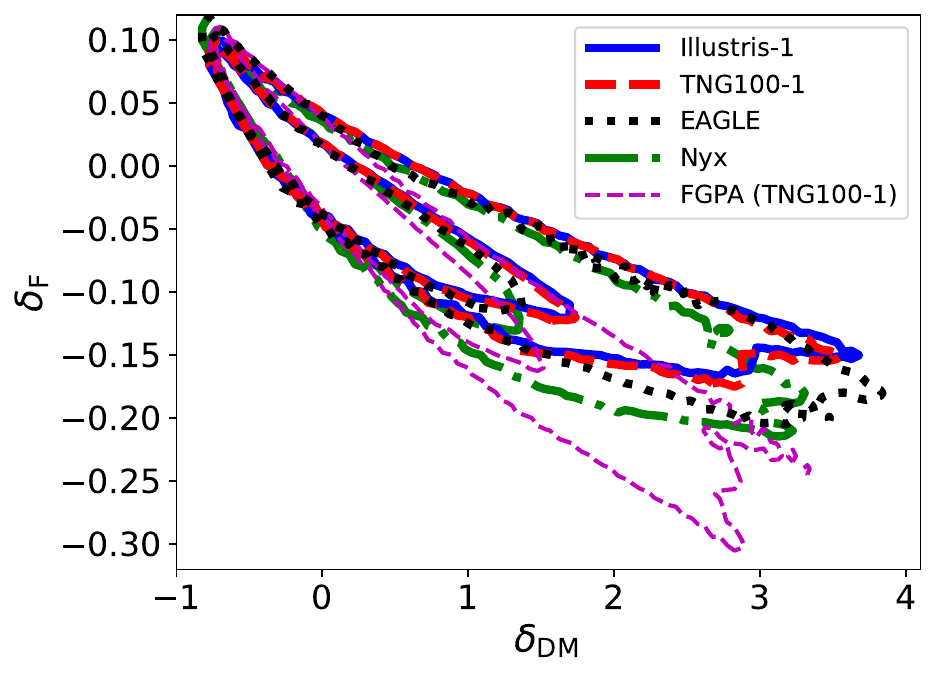}
	\caption{Ly$\alpha$ transmission-DM density distribution for all simulations (thick-lined contours) compared to the FGPA-based distribution from TNG100-1 (thin, solid-lined magenta contours). The contours denote the 2\% and 80\% levels of the PDF of the distribution, respectively. This clearly shows the deviation of the FGPA from the hydrodynamical simulations.}
    \label{fig:lyadm_all}
\end{figure}

Here we can clearly see that at low density ($\delta_{\rm DM} \leq 0$), all the hydrodynamical distributions are consistent with the FGPA distribution. This shows that for underdense regions, the power-law temperature-density relationship provides a good description of the gas properties. However, at higher densities ($\delta_{\rm DM} > 1$), the curves all strongly deviate from the FGPA curve. This is a direct consequence of the power-law temperature-density relationship being a poor representation of the gas at these densities. We note that the DM densities given in the figures have been smoothed to 3 \hmpc{} and one should therefore be aware that the numbers on the axes cannot directly be compared to the values of the unsmoothed density field. Another interesting point to note is that even in the case of Nyx, which includes no feedback prescription, the hydrodynamical distribution deviates from the FGPA. This shows that even in the absence of feedback, baryonic physics beyond simple photoionization equilibrium become significant at those densities, which the collisionless FGPA based skewers cannot model correctly --- this is possibly due to shock heating caused by gravitational collapse of large-scale structure. Any extra feedback mechanisms adopted in the other simulations (Illustris, IllustrisTNG and EAGLE) then tilts the distribution further up towards higher transmission. This is consistent with the findings in \citet{preheatingpaper} in a similar study instead focusing on proto-clusters, where increased levels of cluster pre-heating result in a stronger tilt of the transmission-density distribution.\\

To measure this tilt more quantitatively, we fit a straight line to the distributions for the smoothed density range of $1 \leq \delta_{\rm DM} \leq 3$. The resulting fits are shown as the dotted lines in Figure~\ref{fig:lyadm_indiv} and the values of the slopes are given in the bottom left corner of each panel. Generally, the FGPA contours have a slope around $\alpha\sim-0.08$, whereas the hydrodynamical contours are in the range of $-0.035 \lesssim \alpha \lesssim -0.038$ for simulations with feedback prescriptions. Nyx, which does not model feedback, yields a slightly steeper slope of $\alpha = -0.044$. The FGPA transmission-density relation from Nyx has a slightly steeper slope than the other simulations. This is a consequence of using the baryon velocity field in generating the FGPA. We found that using baryon velocities for the other simulations, similarly steepened the FGPA distribution. Since the DM density field is observationally more easily available, we mainly consider the FGPA curves of EAGLE and Illustris(TNG) when comparing the slopes to the hydrodynamical distribution, but keep the Nyx FGPA curves for completeness.\\

In Figure~\ref{fig:lyadm_all}, we plot the transmission-density relationships from the various hydrodynamical simulations onto the same axes, together with the FGPA distribution based on TNG100-1. There is remarkable agreement between all the distributions at $\delta_{\rm DM}<0$, attesting to the consistency of the photoionization equilibrium physics at low matter densities. Although the differences for the simulations with feedback are small, a weak trend with the strength of the feedback can be seen, where the simulation with no feedback (Nyx) has the steepest slope among the hydrodynamical models, while the strongest feedback (Illustris) yields the flattest slope at high densities. \\

\subsection{Distributions with CLAMATO-like Mock Data}
\begin{figure*}[bt!]\centering
	\includegraphics[width=0.49\columnwidth]{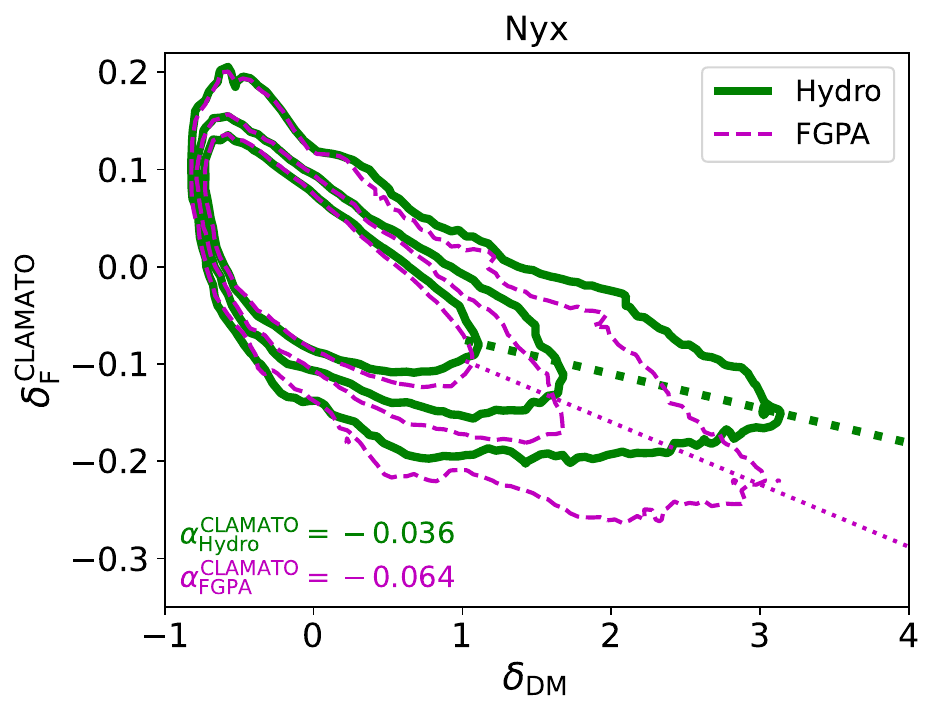} 
	\includegraphics[width=0.49\columnwidth]{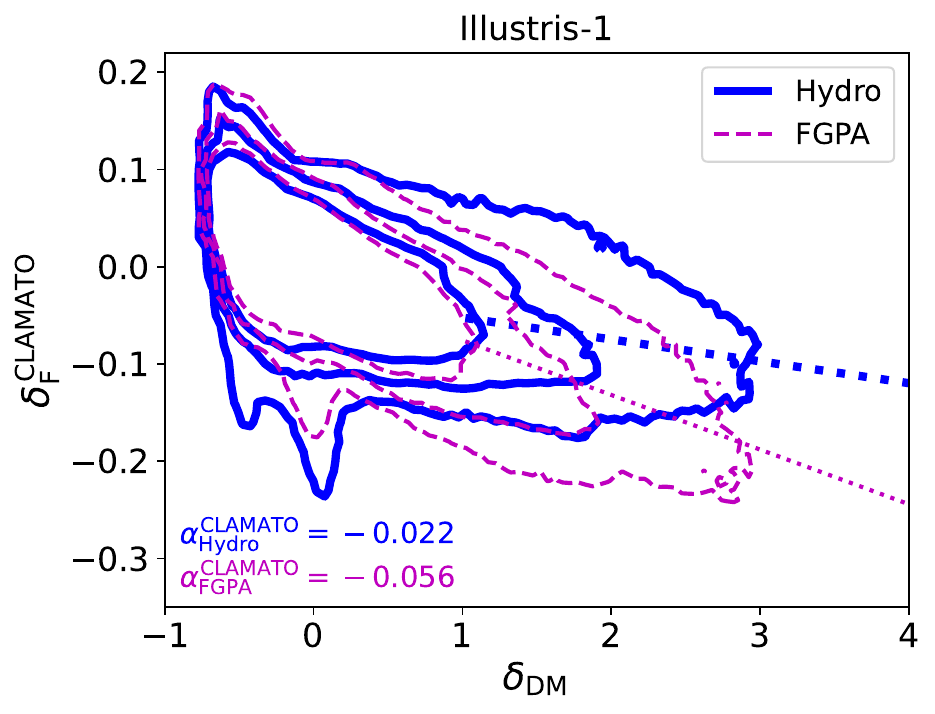}
	\includegraphics[width=0.49\columnwidth]{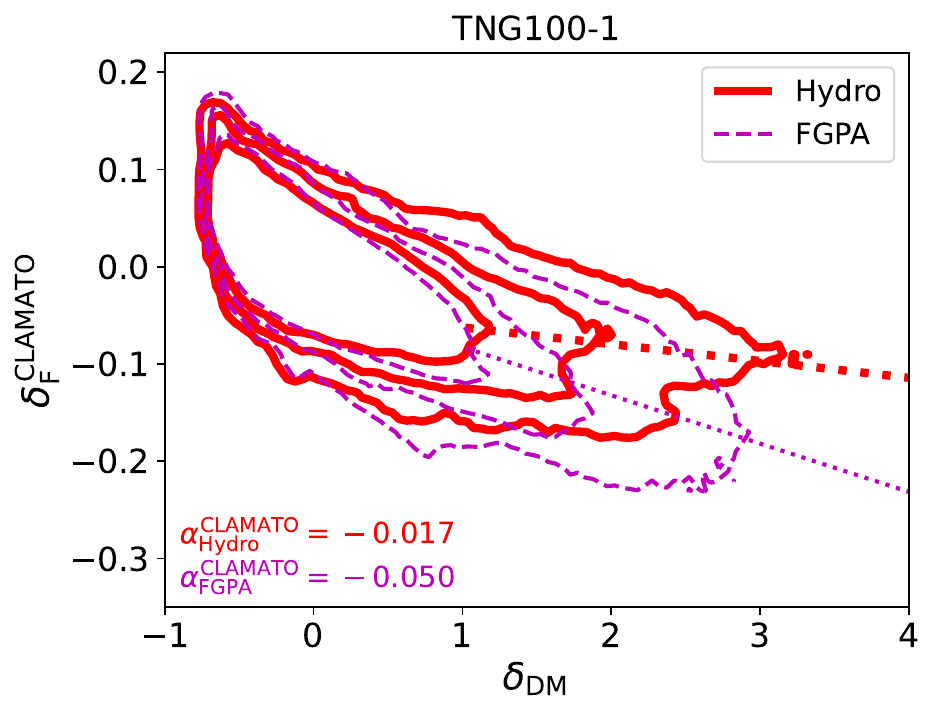} 
	\includegraphics[width=0.49\columnwidth]{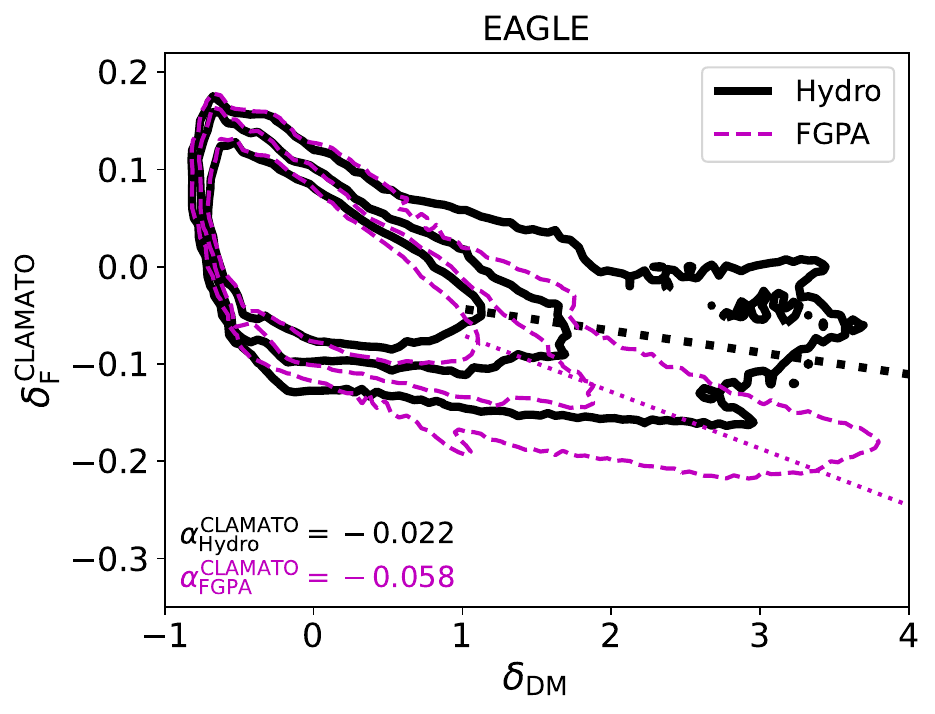}
	\caption{Example realization of the Ly$\alpha$ transmission - DM density distribution of all simulations including CLAMATO-like noise. As before, the contours denote the 2\%, 20\% and 80\% levels of the PDF of the distribution, respectively. In each panel the hydrodynamical skewer distribution is compared to the respective FGPA-based distribution. The dotted lines show the fits to the distribution on the high-density side with the slopes $\alpha$ written in the bottom left corner of every panel, where the thick lines denote the hydrodynamical simulation slopes and the thin lines the FGPA slopes. The difference between the FGPA and hydrodynamical distributions can still clearly be recovered in all cases.}
    \label{fig:lyadm_clamato}
\end{figure*}

The results discussed above were based on the gridded 3D \lya{} transmission, without realistic sightline sampling, noise or spectral smoothing: hence the tight contours. We therefore also generated \lya{} tomographic maps that are more representative of observations such as CLAMATO, the description of which can be found in Section \ref{sec:clamatonoise}. In Figure~\ref{fig:lyadm_clamato}, the reconstructed IGM tomographic maps were also Gaussian smoothed to a scale of $\sigma = 3\,\hmpc$. We show the transmission-density distribution for a single realization of the skewer sampling and noise for
the different simulations. \\

\begin{figure*}[bt]\centering
	\includegraphics[width=0.49\columnwidth]{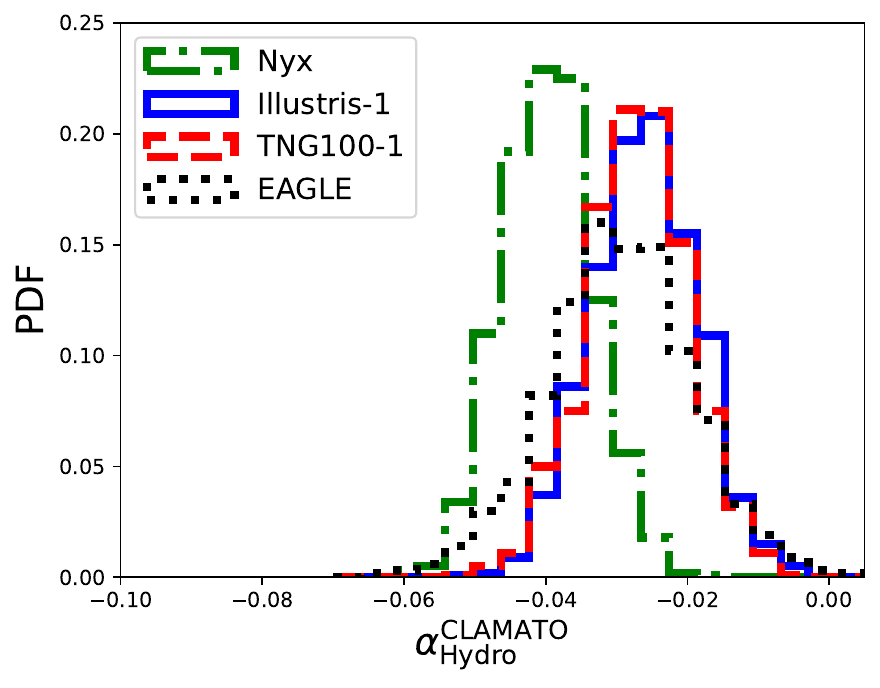} 
\includegraphics[width=0.49\columnwidth]{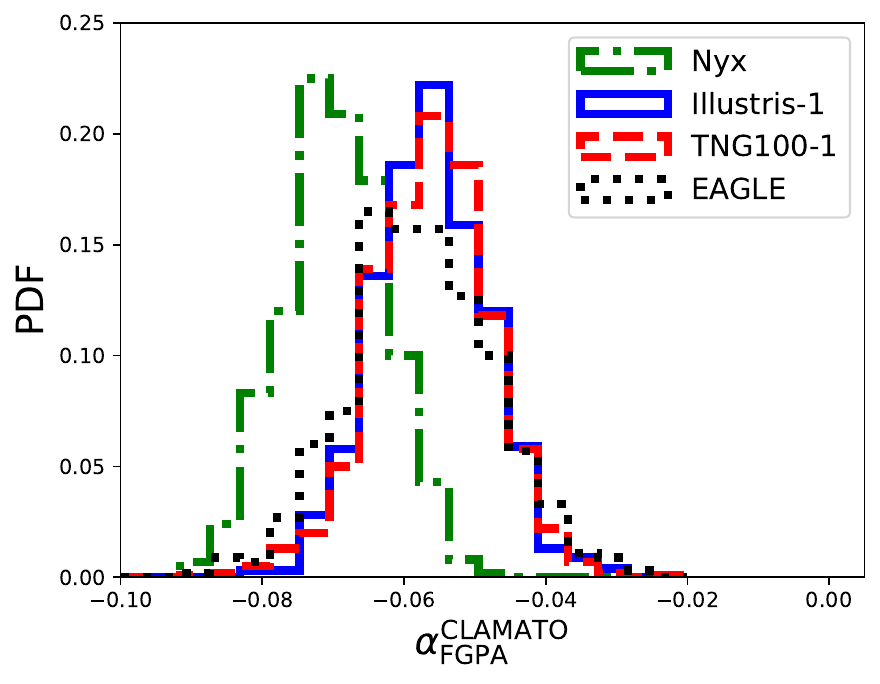}
	\caption{PDFs normalized to unity of the slopes of 1000 realizations of the Ly$\alpha$ transmission - DM density distribution of all simulations including CLAMATO-like noise. The panel on the left shows the distribution for full hydrodynamical skewers, whereas the right panel shows the same for skewers based on the FGPA. The mean of each distribution can be found in Table \ref{tab:slopes}.}
   \label{fig:clamatoslope_pdf}
\end{figure*}

As expected, the instrumental effects result in noisier contours with a much less tight transmission-density relationship than in Section~\ref{sec:sim_dist}. Nevertheless, the difference between the FGPA and the hydrodynamical distributions is still clearly visible in all four simulations, albeit with slopes differing from the noiseless data seen in  Figure~\ref{fig:lyadm_indiv}. Moreover, since these are single realizations of mock CLAMATO-like data sets, the slopes will vary depending on the skewer sampling and noise of the given random realization. To average over these uncertainties, for each simulation or FGPA model we therefore generated 1000 independent realizations and plot the PDFs of the resulting slopes in Figure \ref{fig:clamatoslope_pdf}, where those from the hydrodynamical distribution are shown on the left and the FGPA on the right.\\

Similar to the transmission-density contours without noise, the slope PDFs show a clear separation between the hydrodynamical skewers and those based on the FGPA. The FGPA distributions of the simulations with feedback overlap strongly, while for the Nyx FGPA distribution we again see the slight offset to a steeper slope due to the use of the baryon velocity field to calculate the redshift-space matter density field. On the hydrodynamical side, the distribution of the different realizations' slopes reflect the trends we saw above as well. Non-feedback hydrodynamics in the Nyx simulation already results in an offset from the FGPA due to baryonic physics while the weak trend with the strength of the feedback persists, even in the slope PDFs of the contours including CLAMATO-like noise.\\

\begin{table}[hbt!]
  \centering
	\caption{Measured \lya{} transmission-DM density distribution slopes from the various simulations. The columns give, respectively, the simulation name, the slope of the hydrodynamical \lya{}-DM density distribution without noise, the mean and standard deviation of the hydrodynamical slopes with CLAMATO-like noise, the same with ELT-like noise, the slope of the FGPA transmission-density distribution and the mean and standard deviation of the FGPA slopes with noise for CLAMATO and ELT. The CLAMATO values are based on 1000 realizations, whereas the ELT numbers are based on 100 realizations for EAGLE and 50 realizations for the other simulations.}
	\label{tab:slopes}
    \hspace*{-2cm}
	\begin{tabular}{lcccccc}
		\hline
		Simulation & $\alpha_{\rm Hydro}$ & $\left<\alpha_{\rm Hydro}^{\rm CLAMATO}\right>$ & $\left<\alpha_{\rm Hydro}^{\rm ELT}\right>$ & $\alpha_{\rm FGPA}$ & $\left<\alpha_{\rm FGPA}^{\rm CLAMATO}\right>$ & $\left<\alpha_{\rm FGPA}^{\rm ELT}\right>$\\
		\hline
        Nyx & -0.044 & -0.040 $\pm$ 0.006 & -0.045 $\pm$ 0.002 & -0.087 & -0.070 $\pm$ 0.007 & -0.082 $\pm$ 0.002\\
        Illustris & -0.035 & -0.026 $\pm$ 0.007 & -0.033 $\pm$ 0.002 & -0.080 & -0.056 $\pm$ 0.008 & -0.073 $\pm$ 0.002\\
        IllustrisTNG & -0.036 & -0.027 $\pm$ 0.007 & -0.034 $\pm$ 0.003 & -0.079 & -0.056 $\pm$ 0.008 & -0.073 $\pm$ 0.002\\ 
        EAGLE & -0.040 & -0.030 $\pm$ 0.010 & -0.037 $\pm$ 0.003 & -0.077 & -0.056 $\pm$ 0.010 & -0.071 $\pm$ 0.003\\
		\hline
	\end{tabular}
\end{table}

We summarize the slope values in Table \ref{tab:slopes}, where for every simulation we quote the slope without noise as well as the mean and standard deviation from the ensemble of CLAMATO-like noise reconstructions for the hydrodynamical and FGPA skewers, respectively. We consistently see an offset from the noise-less slope in the means of the slopes with noise towards higher values. This is due to the continuum error adopted for the CLAMATO-like skewers (see Section \ref{sec:clamatonoise}) and such an error would thus have to be taken into account when interpreting the transmission-density distribution from real observations. Furthermore, we see that within 1$\,\sigma$, the FGPA slopes of Illustris, IllustrisTNG and EAGLE are consistent with each other, showing that the transmission-density distribution from the FGPA does not get affected strongly by the different codes, feedback prescriptions, and cosmologies adopted by each simulation, but is instead fully governed by the FGPA slope. The hydrodynamical transmission-density relationship from Nyx is typically $\sim4.1\,\sigma$ removed from its FGPA slopes, whereas for the other simulations the difference varies between $\sim2.8\,\sigma$ to $\sim 4.7\,\sigma$. The differences between the individual simulations with feedback all are smaller than $1\,\sigma$ and thus it would be hard to distinguish between feedback models using a survey such as CLAMATO. The presence of feedback manifests at a low significance around $\sim1\,\sigma$ as seen in the difference between Nyx (which has no feedback) and the other hydrodynamical simulations.\\

\begin{figure*}[b!]\centering
	\includegraphics[width=0.49\columnwidth]{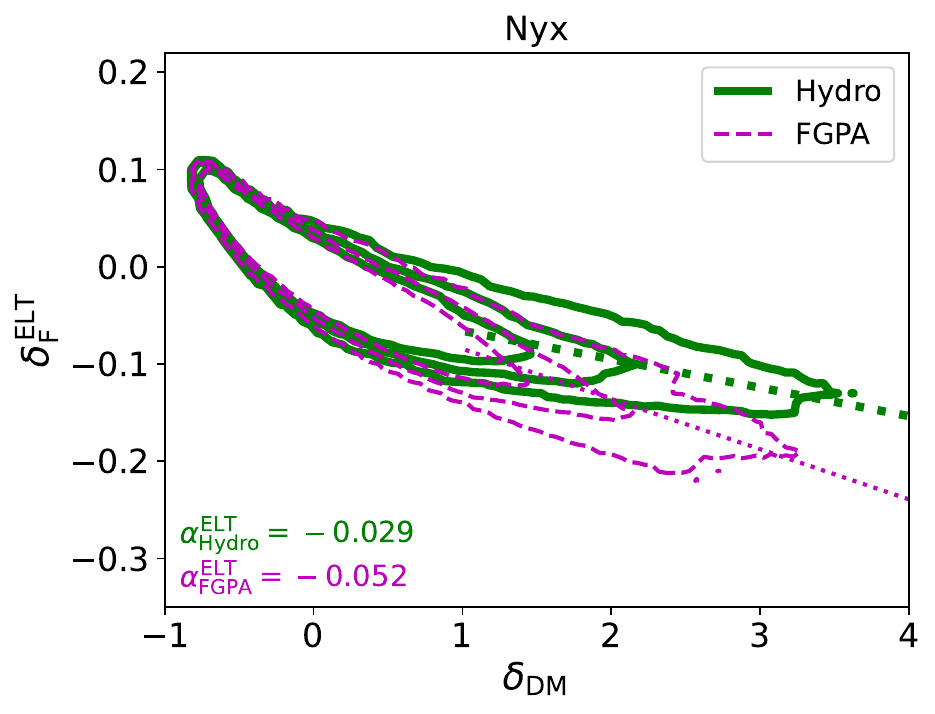} 
	\includegraphics[width=0.49\columnwidth]{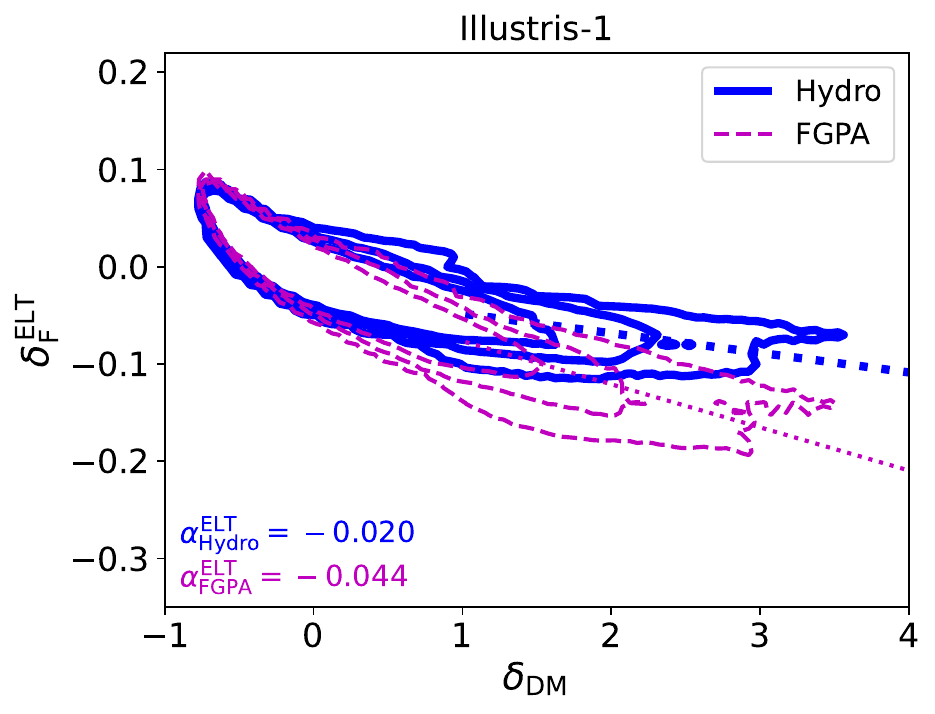}
	\includegraphics[width=0.49\columnwidth]{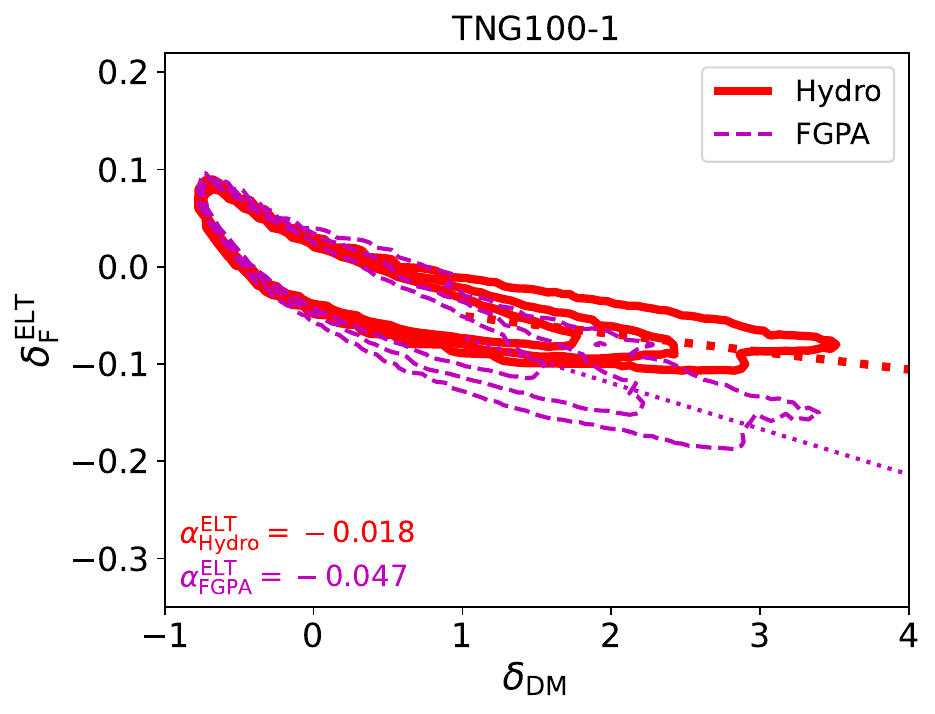} 
	\includegraphics[width=0.49\columnwidth]{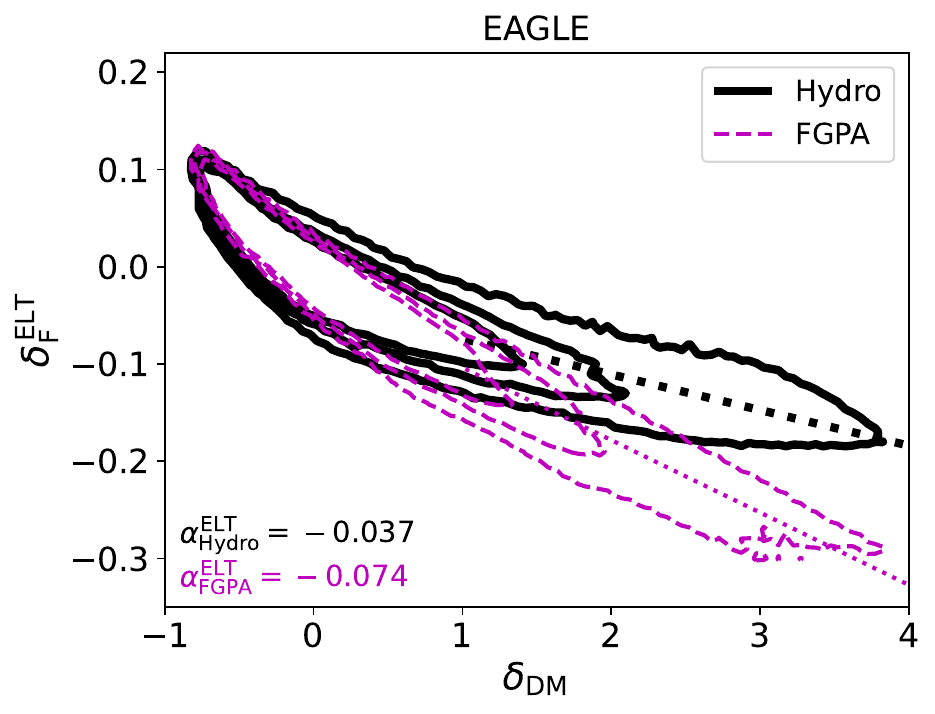}
	\caption{Example realization of the Ly$\alpha$ transmission - DM density distribution of all simulations including ELT-like noise. The contours denote the 2\%, 20\% and 80\% levels of the PDF of the distribution, respectively. In each panel the hydrodynamical skewer distribution is compared to the respective FGPA-based distribution. The dotted lines show the fits to the distribution on the high-density side with the slopes $\alpha$ written in the bottom left corner of every panel, where the thick lines denote the hydrodynamical simulation slopes and the thin lines the FGPA slopes. The difference between the FGPA and hydrodynamical distributions can be recovered for all simulations and the contours are less noisy than was the case for CLAMATO-like realizations in Fig. \ref{fig:lyadm_clamato}.}
    \label{fig:lyadm_elt}
\end{figure*}

Indeed, it would seem that a clear detection of difference between the hydrodynamical and FGPA transmission-density relationships would be marginal using CLAMATO-like data, potentially at the $< 3\sigma$ level. In this analysis, we have only incorporated errors on the IGM tomography data (i.e.\ the $y$-axis in Figure~\ref{fig:lyadm_clamato}), but not from reconstruction uncertainties in the estimated matter density field \citep{birthcosmos}. The errors in the reconstructed matter field would almost certainly introduce further errors into the observed transmission-density relationship, so we do not expect to be able to discriminate between the FGPA and hydrodynamical models with the current generation of observational data. One possibility is to beat down the observational uncertainties with much larger observational volumes: the Subaru PFS Galaxy Evolution Survey (\citealt{nagamine2021}; Greene et al in prep.) would cover $\sim40\times$ greater cosmic volumes than CLAMATO at similar sightline sampling and signal-to-noise, while a coeval galaxy sample is explicitly being designed for the foreground volume mapped by the PFS IGM tomography with comparable galaxy number density as the combined VUDS and zCOSMOS sample at $z\sim 2.3$ \citep{birthcosmos}. This would in principle lead to a $\sim 6-7\times$ reduction in uncertainties in the transmission-density slope, allowing a clear test of the FGPA\footnote{LATIS \citep{latis} already covers nearly an order-of-magnitude larger volume than CLAMATO \citep{qezlou2021}, but it is unclear whether the coeval galaxy sample is sufficient for a matter density reconstruction.}; again, however, a clearer forecast would require modeling of the uncertainties in the density reconstruction, which we will leave for future work.\\

\subsection{Distributions with ELT-like Mock Data}
Observations with future thirty-meter class ELTs are another option to improve the constraints on the FGPA using the transmission-density distribution. Their higher resolution and greater sensitivity will result in more accurate and precise IGM tomography maps that should reduce the scatter in the transmission-density plots compared to a CLAMATO-like survey (e.g.\ in Figure~\ref{fig:lyadm_clamato}), potentially yielding more stringent tests of the FGPA. In order to quantify this improvement, we ran another set of mock realizations based on the survey parameters that should be feasible with the ELTs (see Section \ref{sec:clamatonoise}).\\

\begin{figure*}[bt!]\centering
	\includegraphics[width=0.49\columnwidth]{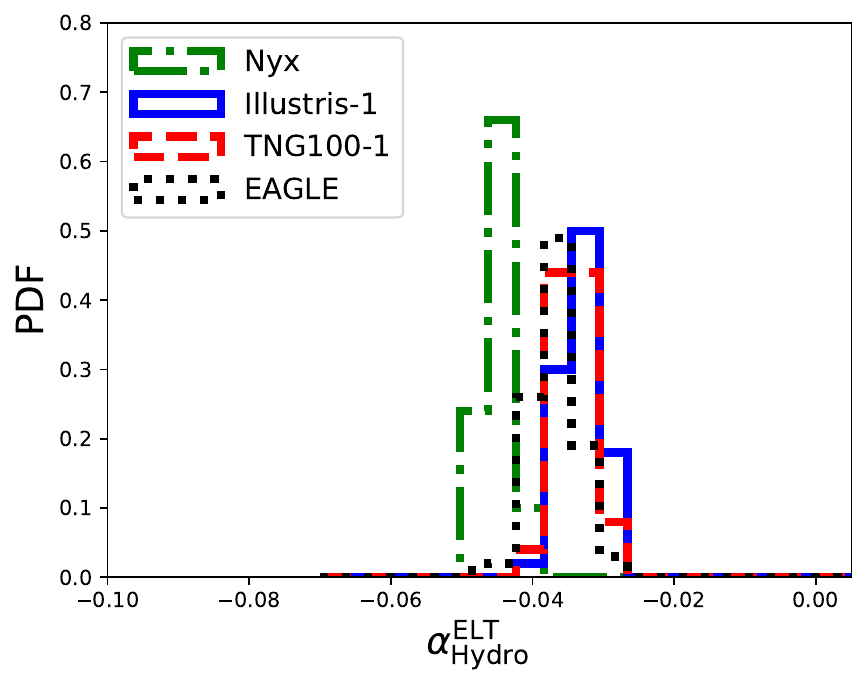} 
\includegraphics[width=0.49\columnwidth]{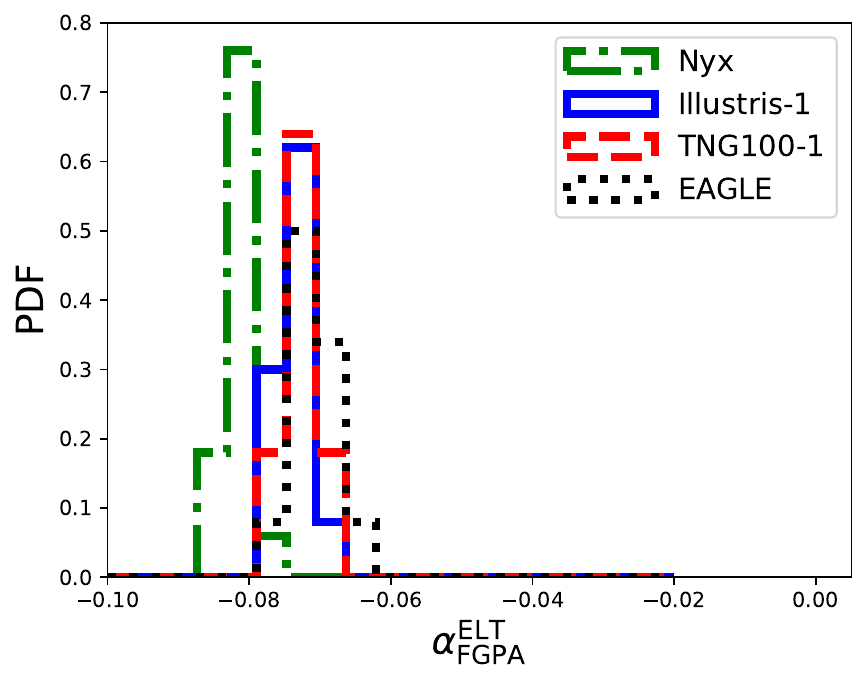}
	\caption{PDFs normalized to unity of the slopes of the Ly$\alpha$ transmission - DM density distribution of all simulations including ELT-like noise. 100 realizations were used for EAGLE and 50 realizations for the other simulations. The panel on the left shows the distribution for full hydrodynamical skewers, whereas the right panel shows the same for skewers based on the FGPA. The mean of each distribution can be found in Table \ref{tab:slopes}.}
   \label{fig:eltslope_pdf}
\end{figure*}

Due to the higher resolution and larger numbers of skewers for every simulation, the computational time required for the Wiener-filtering increased dramatically with the higher sightline density. Therefore, we computed 100 realizations each of mock hydrodynamical skewers and FGPA skewers for the EAGLE simulation and only 50 realizations each for Illustris, IllustrisTNG and Nyx. Example realizations of ELT-like noise transmission-density distributions are shown in Fig. \ref{fig:lyadm_elt}. Following expectations, the contours in this case are much less noisy than was the case for CLAMATO observations in Fig. \ref{fig:lyadm_clamato}. Similar to the CLAMATO distributions there, ELT-like noise also in all cases yields a clear difference between the hydrodynamical distribution and the FGPA.\\

The normalized PDFs of all the realizations can be found in Fig. \ref{fig:eltslope_pdf} and the mean values in Table \ref{tab:slopes}. Due to the better specifications of ELTs compared to CLAMATO, the PDFs are also narrower, now clearly showing the separation between FGPA and hydrodynamical skewers. The mean values are also closer to the noiseless distributions than was the case with CLAMATO-like noise, while the difference between the mean hydrodynamical slope and mean FGPA slope has now increased to $\sim$12-20$\sigma$. The difference between simulations including feedback and Nyx has consequently also increased to $\sim$3.3-6$\sigma$. This shows that ELT-class telescopes will significantly improve the measurements proposed in this work: it should clearly detect the deviation from the FGPA in the transmission-density relationship, as well as potentially be able to detect the effect of feedback on the IGM at Cosmic Noon (i.e.\ the difference between Nyx and the other hydro simulations). We note, that the already small $\lesssim1\sigma$ differences between the individual feedback simulations (Illustris, IllustrisTNG and EAGLE) are too small to measure in the ELT case as well. As was mentioned above, more work would be needed to incorporate the uncertainties in the reconstructed matter field to yield a more quantitative forecast, although we can also expect reduced uncertainties on this axis from improvements in observational sampling as well as better methodology. We will again leave this analysis to future work. 

\begin{figure*}[b!]\centering
	\includegraphics[width=0.49\columnwidth]{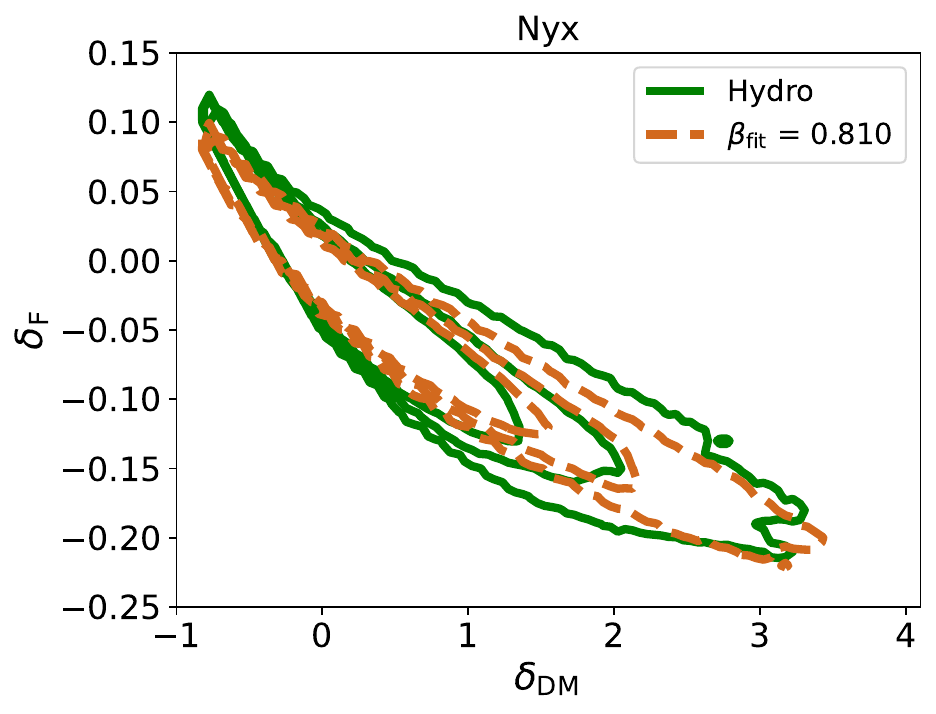} 
	\includegraphics[width=0.49\columnwidth]{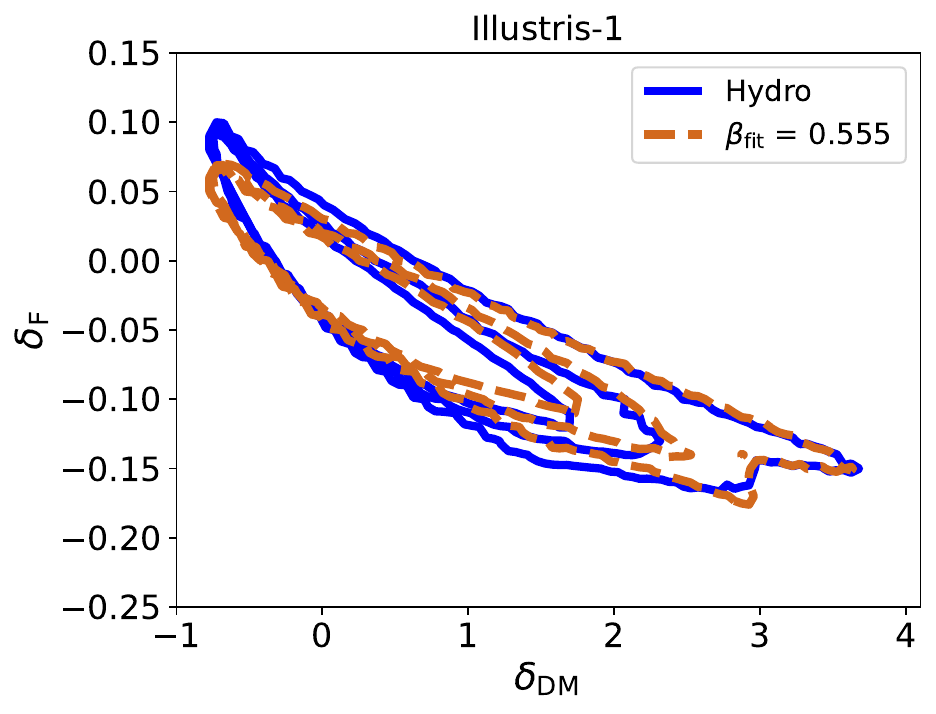}
	\includegraphics[width=0.49\columnwidth]{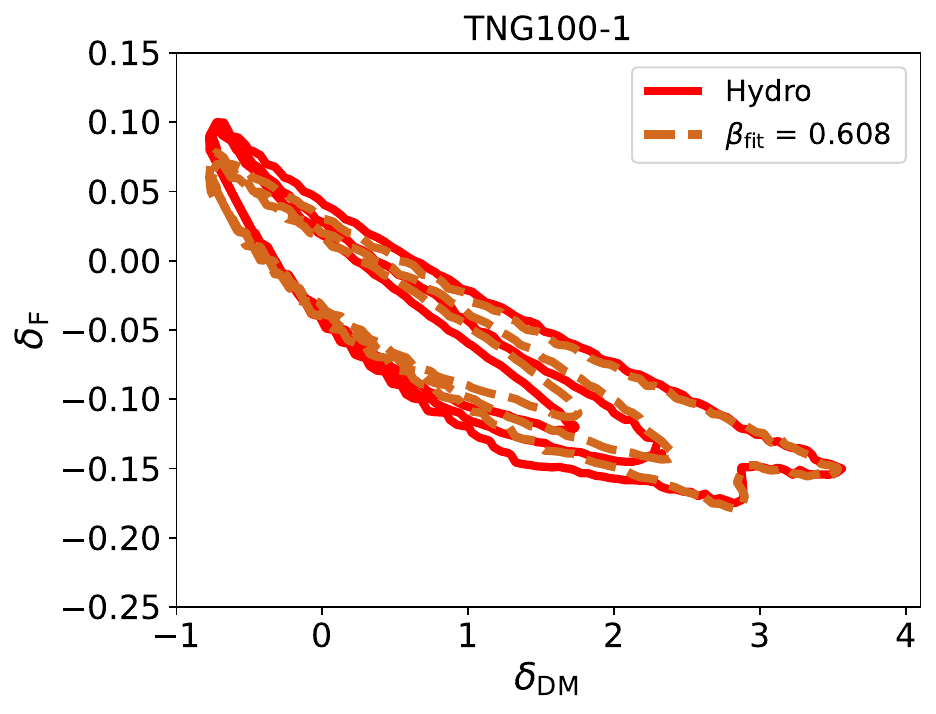} 
	\includegraphics[width=0.49\columnwidth]{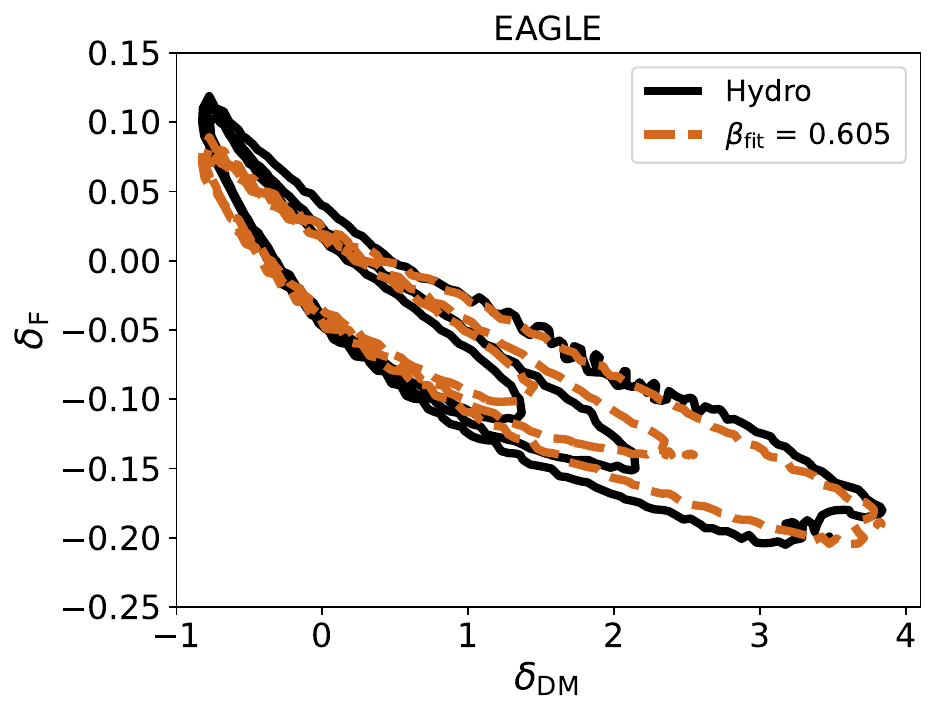}
	\caption{Ly$\alpha$ transmission-DM density distribution for all simulations compared to the FGPA-based distribution with the best-fit $\beta$ slope. The contours again denote the 2\%, 20\% and 80\% levels of the PDF of the distribution, respectively. The best-fit FGPA slope values are given in the legends of each respective panel.}
    \label{fig:lyadm_bestfits}
\end{figure*}

\subsection{Fitting FGPA}\label{sec:fgpafit}
So far, our implementation of the FGPA assumes the canonical slope of $\beta = 1.6$ that arises from adopting the gas temperature-density relationship of $\gamma=1.6$. This, as we have seen, causes the transmission-density relationship to 
deviate at regions greater than the cosmic mean density.
To alleviate this, we explored ways of modifying the FGPA to get a better match with the hydrodynamical skewers. First, we set up a grid of FGPA transmission-density distributions by varying the value of the FGPA slope $\beta$. How a variation of this slope affects the resulting distribution can be seen in Figures \ref{fig:fitgrid_nyx}, \ref{fig:fitgrid_ill}, \ref{fig:fitgrid_tng} and \ref{fig:fitgrid_eagle} in Appendix \ref{app:fitgrids}. From these grids it is clear that the FGPA slope needs to be smaller than the canonical value based on $\beta=1.6$, in order to better reproduce the high-density side of the transmission-density distribution.\\

We therefore created a 3D interpolation function which interpolates the full 2D shape of the transmission-density distribution as a function of the FGPA slope $\beta$. We applied this function to each of the simulations separately, yielding the best-fit $\beta$-slope values and corresponding FPGA transmission-density distributions shown in Figure~\ref{fig:lyadm_bestfits}.\\

Although the correspondence on the low-density end, which was well represented by the canonical FGPA, is degraded with these fitted slope values, the rest of the distribution is much better matched. Therefore, these $\beta$-values could be adopted to update the FGPA for DM-only simulations to reproduce the transmission-density distribution based on any of the feedback models of the simulations adopted in this study. However, the FGPA using these new power-law indices would need to be tested carefully against other \lyaf{} statistics such as the 1D transmission power spectrum and 3D transmission power spectrum before it can be considered robust for general usage. We show the former in Figure~\ref{fig:pspecs} in the Appendix and see that the amplitude is consistently lower than that of the hydrodynamical skewers. Moreover, the transmission PDF (see Figure~\ref{fig:fluxpdfs} in the Appendix) is also highly skewed for the skewers with the fitted FGPA slopes. \\

Since the differences in the FGPA slope seems to be more relevant for the high-density side of the distribution, we attempted to create mock skewers with a double power-law. Regions with low-density would receive an FGPA slope corresponding to the canonical $\gamma=1.6$ of the temperature-density relationship, whereas high-density regions would adopt the slope we found here in this work. The cut-off value for the change to high-density was manually adjusted. We found it difficult to get such a double power-law distribution to properly match the hydrodynamical distribution. The more abundant low-density cells with the canonical FGPA slope will get blended with the less abundant high-density cells due to the smoothing of the final maps. As a consequence the resulting distribution would often be closer to the standard FGPA distribution than the distribution based on the fitted value. We therefore suggest reworking the fitting procedure we presented here by leaving the slopes on both density sides, as well as the cut-off value as a free parameter. Since this would significantly increase the computational expense, we leave this for future work.

\section{Conclusions}

The fluctuating Gunn-Peterson approximation is commonly-used to generate mock \lyaf{} skewers from collisionless DM-only simulations. Using a collection of four hydrodynamical simulations at $z=2$ and their underlying matter-density distributions, we tested the validity of the FGPA in the context of the \lya{} transmission-DM density distribution. We find that the slope of this distribution can be sensitive to the processes governing the heating and cooling of the IGM in each simulation, although the effect on the \lya{} transmission-DM density distribution is small.\\

Matter overdensities ($\delta_\mathrm{DM}>0$) generally exhibit more sensitivity toward the feedback models and we find a deviation from the FGPA in the transmission-density slopes in all the hydrodynamical simulations considered here. This includes Nyx, a hydrodynamical simulation without stellar or AGN feedback, where nevertheless baryonic effects from nascent non-linear structure formation cause a flattening of the transmission-density curve compared to the FGPA. For models that do include a feedback prescription (i.e.\ Illustris, IllustrisTNG and EAGLE), we find that the distribution flattens out even further towards larger \lya{} transmission at high densities. At low densities, we find that the FGPA provides a good description of the gas properties, yielding a good match to the transmission-density distribution seen in hydrodynamical simulations. We expect that at higher redshifts than the $z=2$ studied here, the validity regime of the FGPA might improve toward higher matter densities since non-linear structure formation and feedback have had less of an imprint on the IGM, but we leave this to future work. \\

We next proceeded to study whether the transmission-density distribution from a realistic observational survey volume could place constraints on the FGPA, by generating \lya{} tomographic maps with noise and the resolution similar to an IGM tomographic survey such as CLAMATO. We find a marginal difference $\sim3-4\sigma$ between the FGPA and the hydrodynamical models that would be challenging to recover in existing observations of CLAMATO, in conjunction with the matched density reconstructions from galaxy redshifts in the coeval volume. Upcoming large surveys such as Subaru PFS, however, could cover enough cosmic volume to mitigate the observational uncertainties and firm tests of the FGPA.
With the IGM tomographic observations feasible with upcoming ELTs, however, the significantly improved sightline sampling and spectral signal-to-noise should allow deviations from the FGPA to be discerned from the observed transmission-density relationships with as much as $\sim20\,\sigma$ within observational footprints of $\sim 1\,\mathrm{deg}^2$. Additionally, the difference between any feedback and no feedback (i.e. Nyx) can be recovered with a significance up to $\sim3.6\sigma$. We also find weak evidence for a trend in the slope of the transmission-density distribution with the strength of the heating in the feedback model, where the distribution from Illustris, which includes the strongest feedback, results in the steepest slope. However, we note that the significance of this trend is less than $1\,\sigma$ and would likely not be discernable from the global transmission-density relationship even with future surveys, although focusing on galaxy protoclusters might yield clearer constraints \citep{preheatingpaper}.\\

Finally, we provide fit values for the FGPA slope of each simulation that best reproduces the transmission-density distribution. This could be beneficial for studies focusing on constraining this distribution and will allow for studies using large-volume DM-only simulations to generate mock \lyaf{} skewers with an FGPA slope that matches the desired simulation model. These fits could also inform hydrodynamical models used in forward modelling frameworks for \lyaf{} tomographic reconstructions \citep{tardis1,tardis2}.\\

Upcoming spectroscopic surveys targeting the epoch of Cosmic Noon, such as those using Subaru PFS and thirty meter-class telescopes will potentially be beneficial to increase the statistical power of this technique and help to constrain AGN and stellar feedback models in a novel way.

\begin{acknowledgments}
The authors would like to express their gratitude to Ilya Khrykin, Metin Ata, Joe Hennawi, Davide Martizzi and Daniele Sorini for their useful discussions and input throughout various stages of the project. KGL acknowledges support from JSPS Kakenhi Grants JP18H05868 and JP19K14755.
\end{acknowledgments}

%





\bibliography{FGPA}{}
\bibliographystyle{aasjournal}


\appendix

\section{Lyman alpha statistics}\label{app:lyastat}
This appendix provides some of the common \lyaf{} statistics derived for each simulation following the descriptions in Section \ref{sec:tomogr}. 

\subsection{Single skewers}\label{app:singskew}
In the figures below we show the properties along a single 1D skewer of each simulation. These include the \lya{} transmission based on the full hydrodynamical and the FGPA calculations, the baryon and DM density, the baryon and DM velocity along the sight line and the gas temperature. As can be seen, the DM-based FGPA skewers roughly recover the same absorption features as the hydrodynamical ones, but the details such as the strength and width of the absorption vary.

\begin{figure}[hbt!]
    \centering
	\includegraphics[width=0.73\columnwidth]{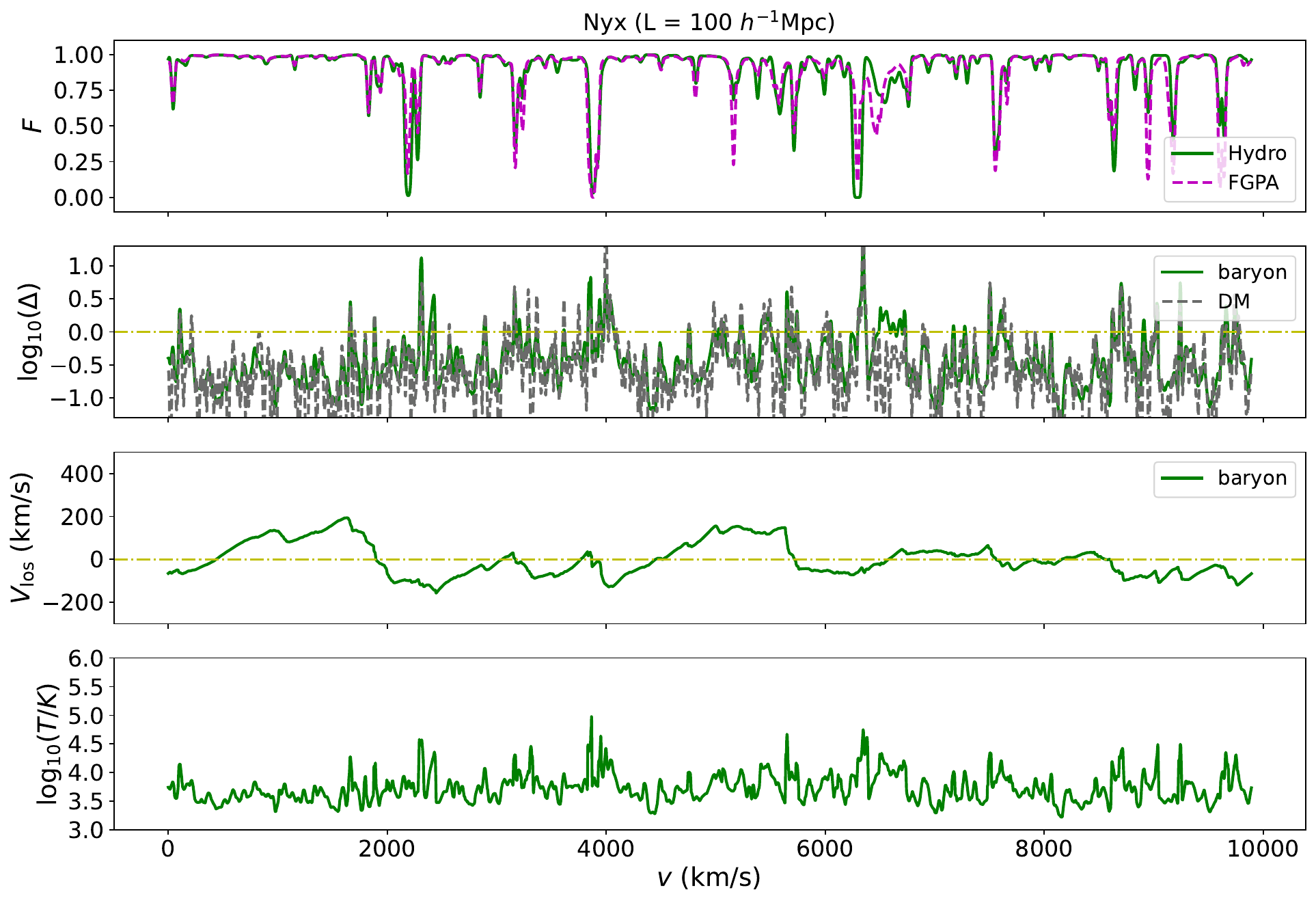}
	\caption{Properties of a single skewer through the center of the Nyx simulation. The top panel shows the Ly$\alpha$ transmission of the hydrodynamical skewer, as well as the same sightline from the FGPA skewers. The second panel displays the density along the line. The third panel shows the velocity along the skewer. We note that the Nyx output does not contain DM velocities. The final panel gives the gas temperature along the skewer.}
    \label{fig:sing_nyx}
\end{figure}
\clearpage

\begin{figure}[hbt!]
    \centering
	\includegraphics[width=0.73\columnwidth]{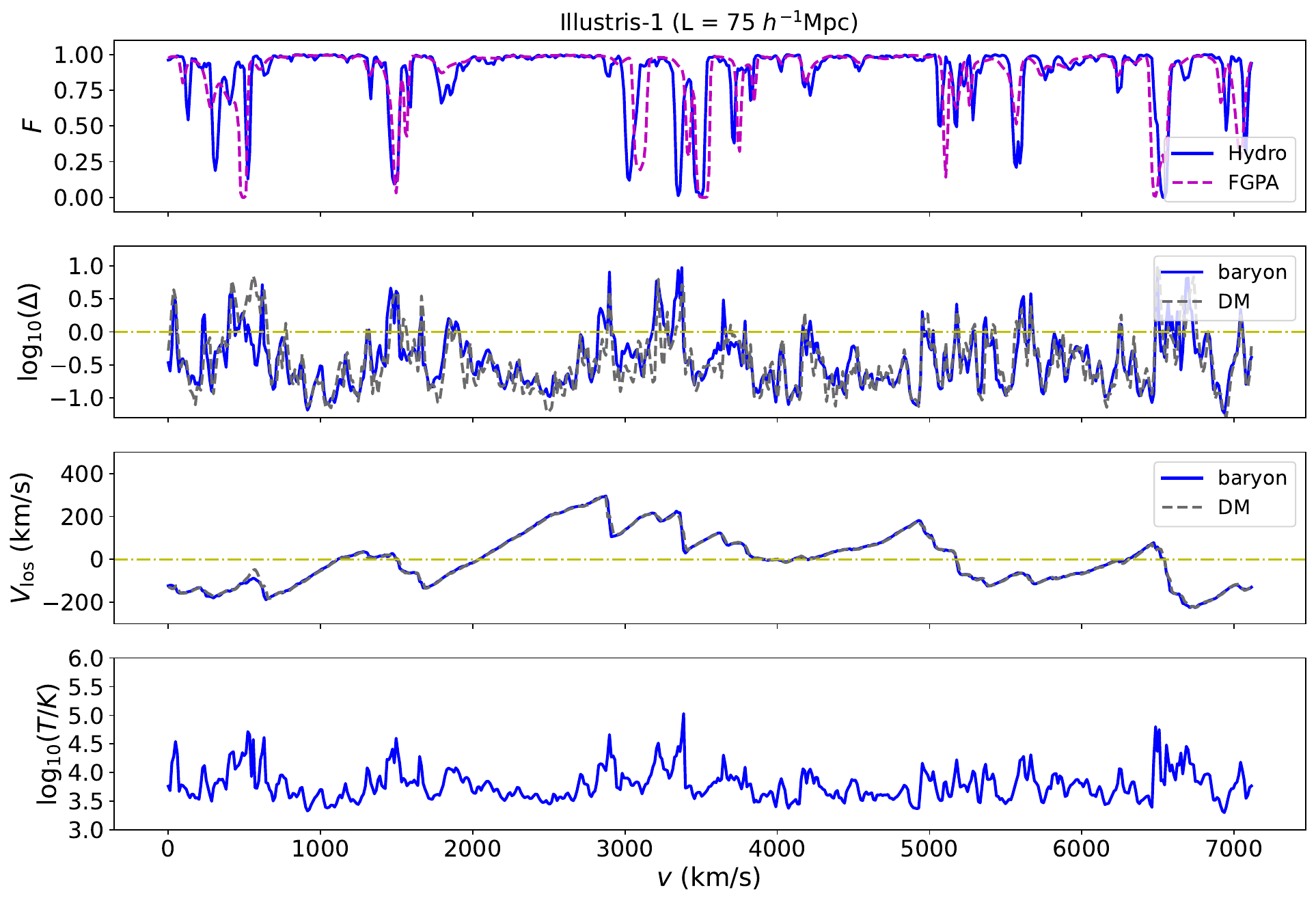}
	\caption{Properties of a single skewer through the center of the Illustris-1 simulation. The top panel shows the Ly$\alpha$ transmission of the hydrodynamical skewer, as well as the same sightline from the FGPA skewers. The second panel displays the density along the line. The third panel shows the velocity along the skewer. The final panel gives the gas temperature along the skewer.}
    \label{fig:sing_ill}
\end{figure}

\begin{figure}[hbt!]
    \centering
	\includegraphics[width=0.73\columnwidth]{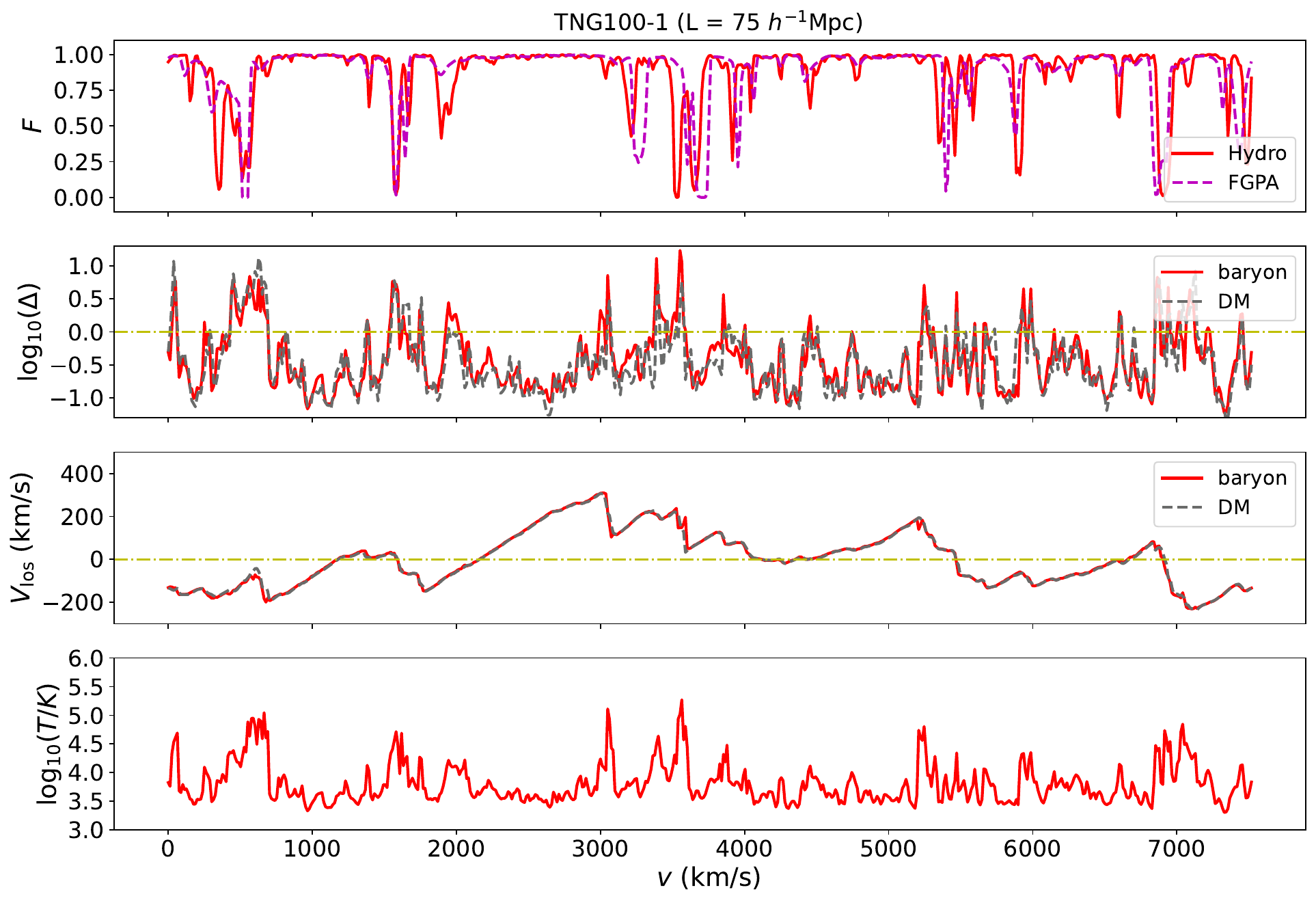}
	\caption{Properties of a single skewer through the center of the TNG100-1 simulation. The top panel shows the Ly$\alpha$ transmission of the hydrodynamical skewer, as well as the same sightline from the FGPA skewers. The second panel displays the density along the line. The third panel shows the velocity along the skewer. The final panel gives the gas temperature along the skewer.}
    \label{fig:sing_tng}
\end{figure}

\clearpage
\begin{figure}[hbt!]
    \centering
	\includegraphics[width=0.73\columnwidth]{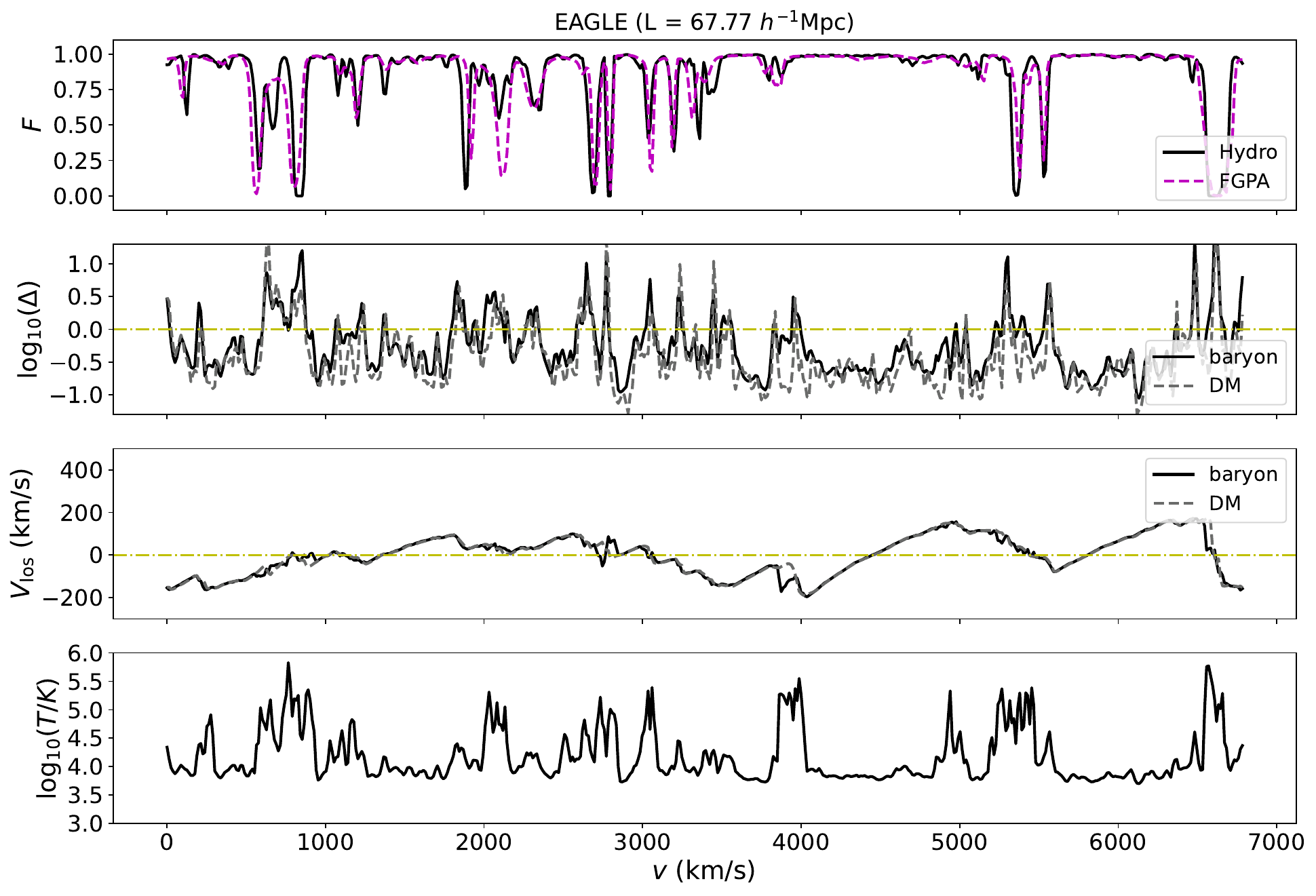}
	\caption{Properties of a single skewer through the center of the EAGLE simulation. The top panel shows the Ly$\alpha$ transmission of the hydrodynamical skewer, as well as the same sightline from the FGPA skewers. The second panel displays the density along the line. The third panel shows the velocity along the skewer. The final panel gives the gas temperature along the skewer.}
    \label{fig:sing_eagle}
\end{figure}

\subsection{Transmission PDF}\label{app:fluxpdf}
Below we present the PDFs of the \lya{} transmission taken directly from both the hydrodynamical and FGPA-based \lya{} tomographic maps of every simulation. Also shown are the distributions based on the skewers with the best-fit FGPA slope value following Section \ref{sec:fgpafit}, as well as the observational data at $z$ = 2.07 by \citet{kim2007}. The PDFs of the FGPA slope fits generally provide a poor match to the observed flux PDF.

\clearpage

\begin{figure*}[hbt!]\centering
	\includegraphics[width=0.49\columnwidth]{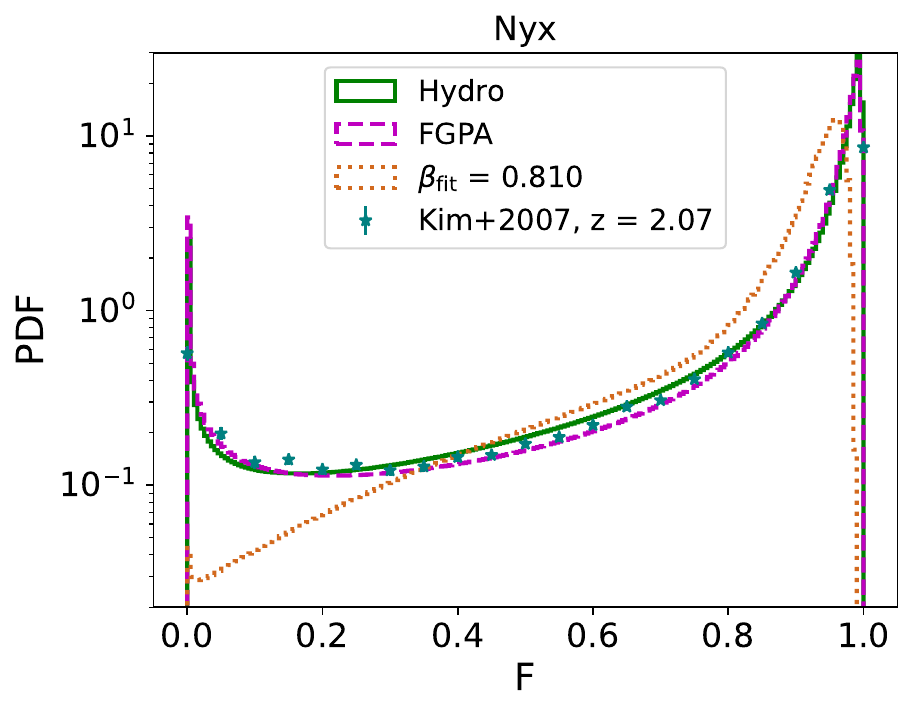} 
	\includegraphics[width=0.49\columnwidth]{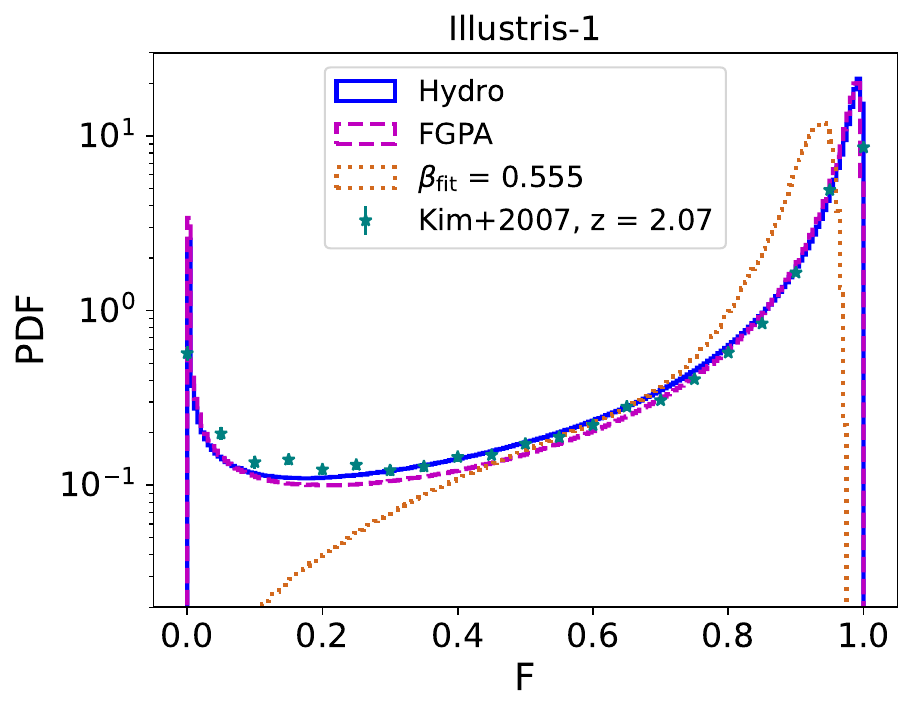}
	\includegraphics[width=0.49\columnwidth]{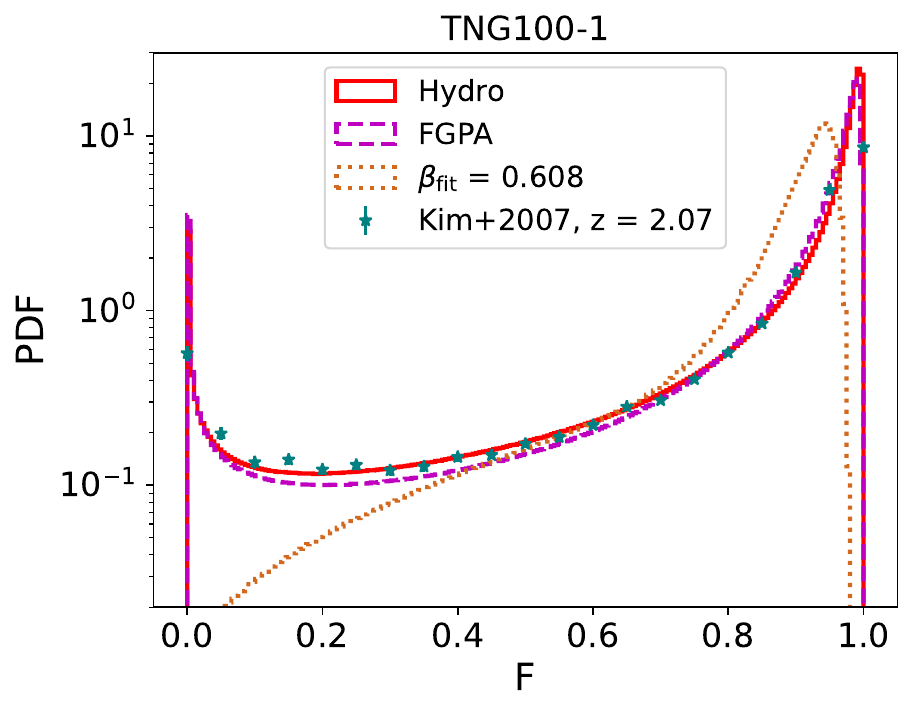} 
	\includegraphics[width=0.49\columnwidth]{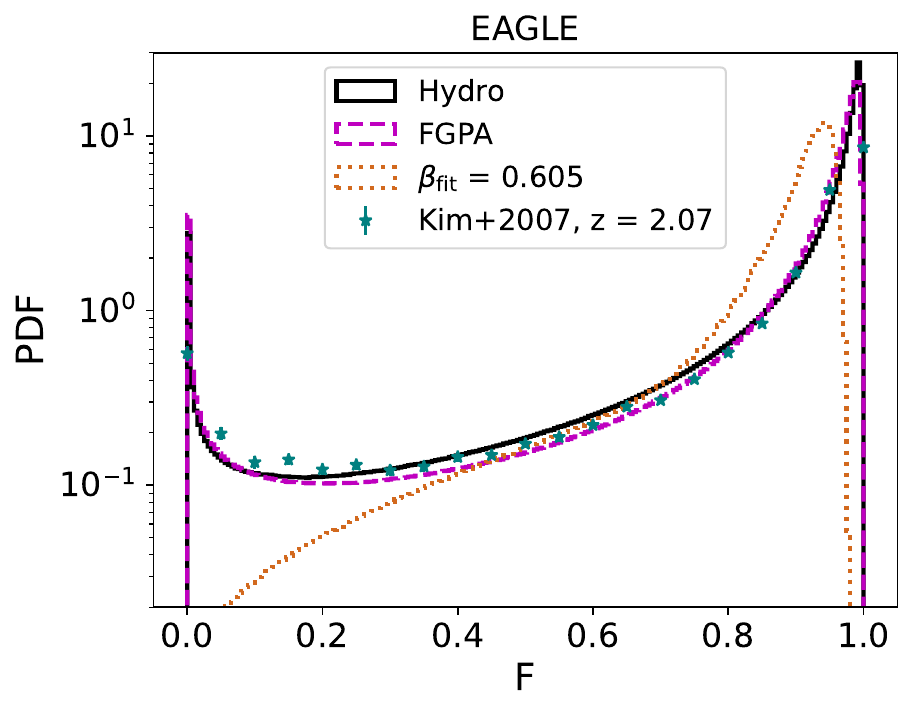}
	\caption{Probability distribution functions of the Ly$\alpha$ transmission. For each simulation we compare the distribution of the hydrodynamical skewers (solid lines) to that of the FGPA-based skewers (dashed lines) and the FGPA-based skewers with best-fit slope values (dotted lines). The observed datapoints by \citet{kim2007} are also shown for reference (stars).}
    \label{fig:fluxpdfs}
\end{figure*}

\subsection{Power spectra}\label{app:pspec}
\begin{figure*}[hbt!]\centering
	\includegraphics[width=0.49\columnwidth]{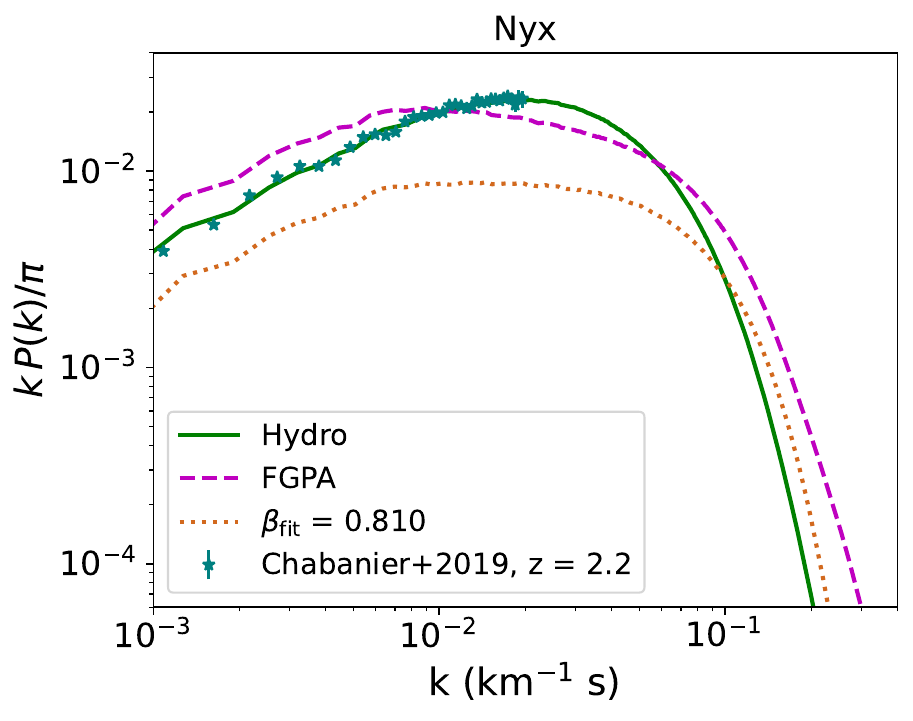} 
	\includegraphics[width=0.49\columnwidth]{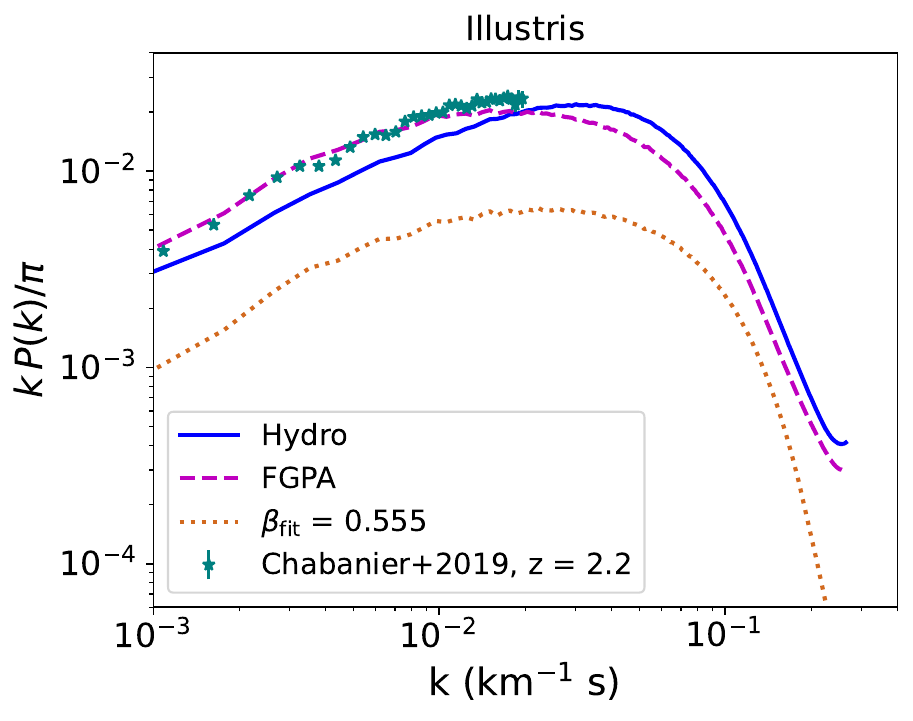}
	\includegraphics[width=0.49\columnwidth]{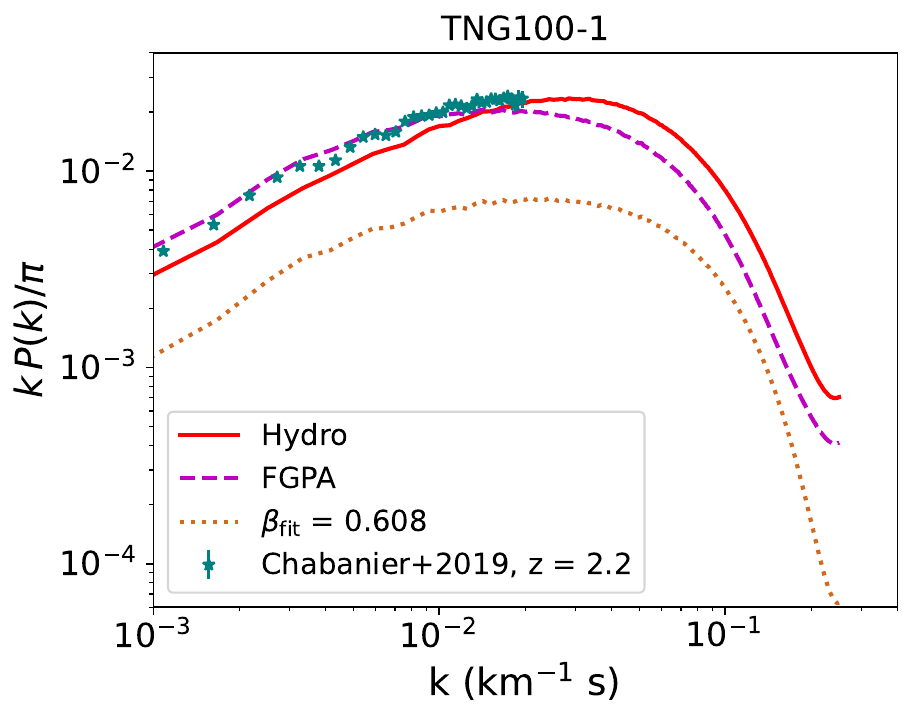} 
	\includegraphics[width=0.49\columnwidth]{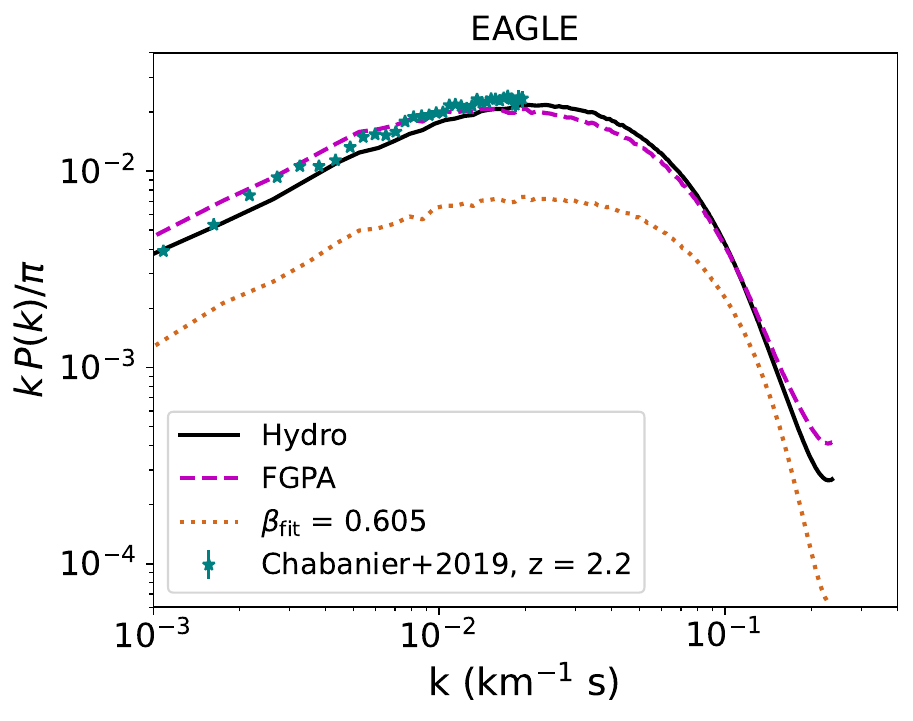}
	\caption{1D line power spectra of the hydrodynamical skewers of the different simulations (solid lines) compared to the power spectra for the FGPA-based skewers (dashed lines) and the FGPA-based skewers with best-fit slope values (dotted lines). The stars denote the observed power spectrum at $z$ = 2.2 by \citet{Chabanier2019}, which have been down-scaled by a factor of 1.7 to match the amplitude of the Nyx power spectrum at $z$ = 2.}
    \label{fig:pspecs}
\end{figure*}

The combined 1D line power spectrum $P(K)$ of multiple \lyaf{} skewers is defined as the follows:
\begin{equation}
P(k) = \left<\left|\bar{\delta_{\rm F}}(k)\right|^2\right>_{N_{\rm skew}},
\end{equation}
where $\bar{\delta_{\rm F}}(k)$ is the Fourier transform of the transmission overdensity at wavemode $k$ and the brackets denote the ensemble average over $N_{\rm skew}$ skewers. We show the resulting unit-less power spectra $kP(k)/\pi$ in Figure~\ref{fig:pspecs}.

\clearpage

\section{FGPA fitting grids}\label{app:fitgrids}
The figures below present the grids along the FGPA slope $\beta$ that were used to fit the best FGPA slope to the hydrodynamical distribution for each simulation in Section \ref{sec:fgpafit}. These clearly show that the distribution tilts up towards higher transmission as $\beta$ decreases.

\begin{figure}[hbt!]
    \centering
	\includegraphics[width=\columnwidth]{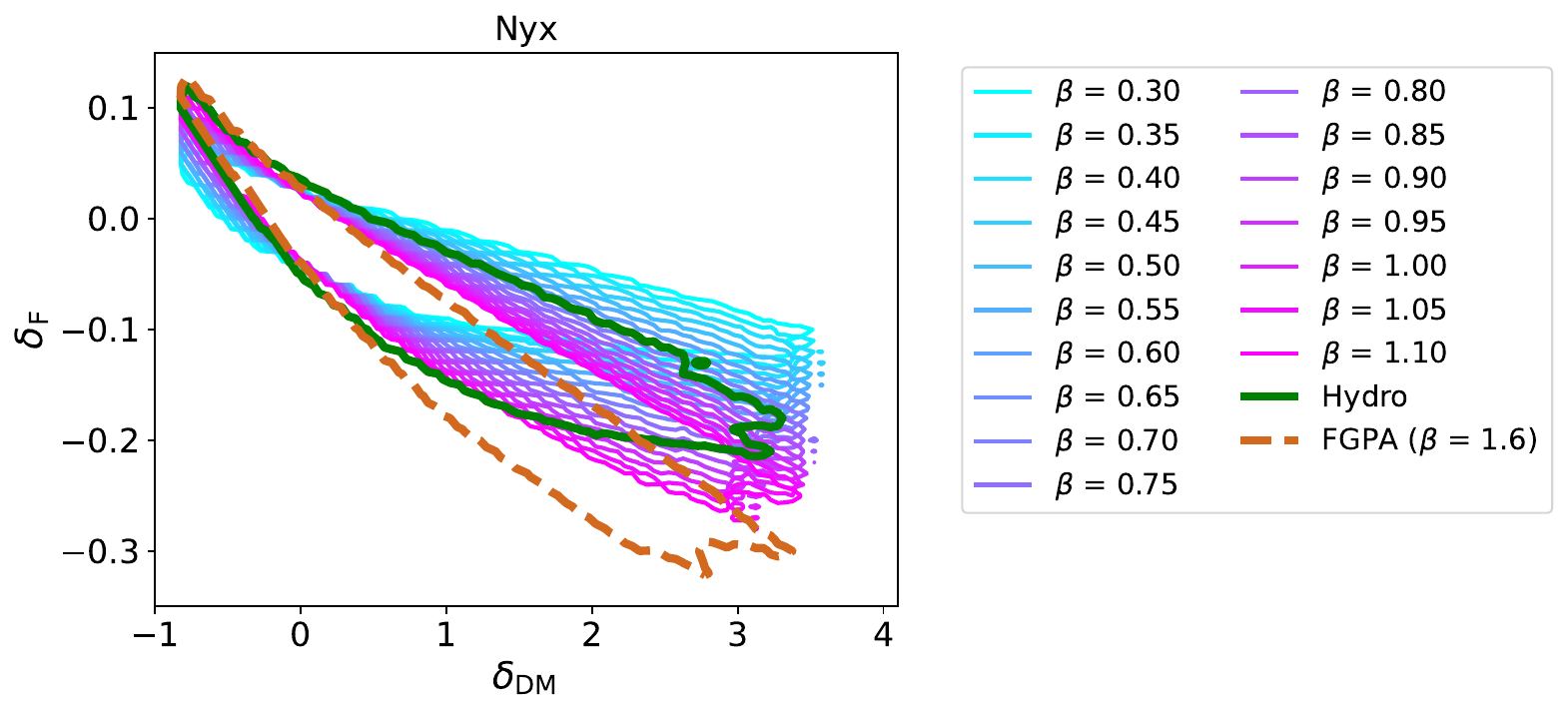}
	\caption{Ly$\alpha$ transmission-DM density distribution for Nyx compared to the FGPA-based distribution as well as the grid with varying FGPA $\beta$ slopes. The contours show the 2\% level of the PDF of the distribution.}
    \label{fig:fitgrid_nyx}
\end{figure}

\begin{figure}[hbt!]
    \centering
	\includegraphics[width=\columnwidth]{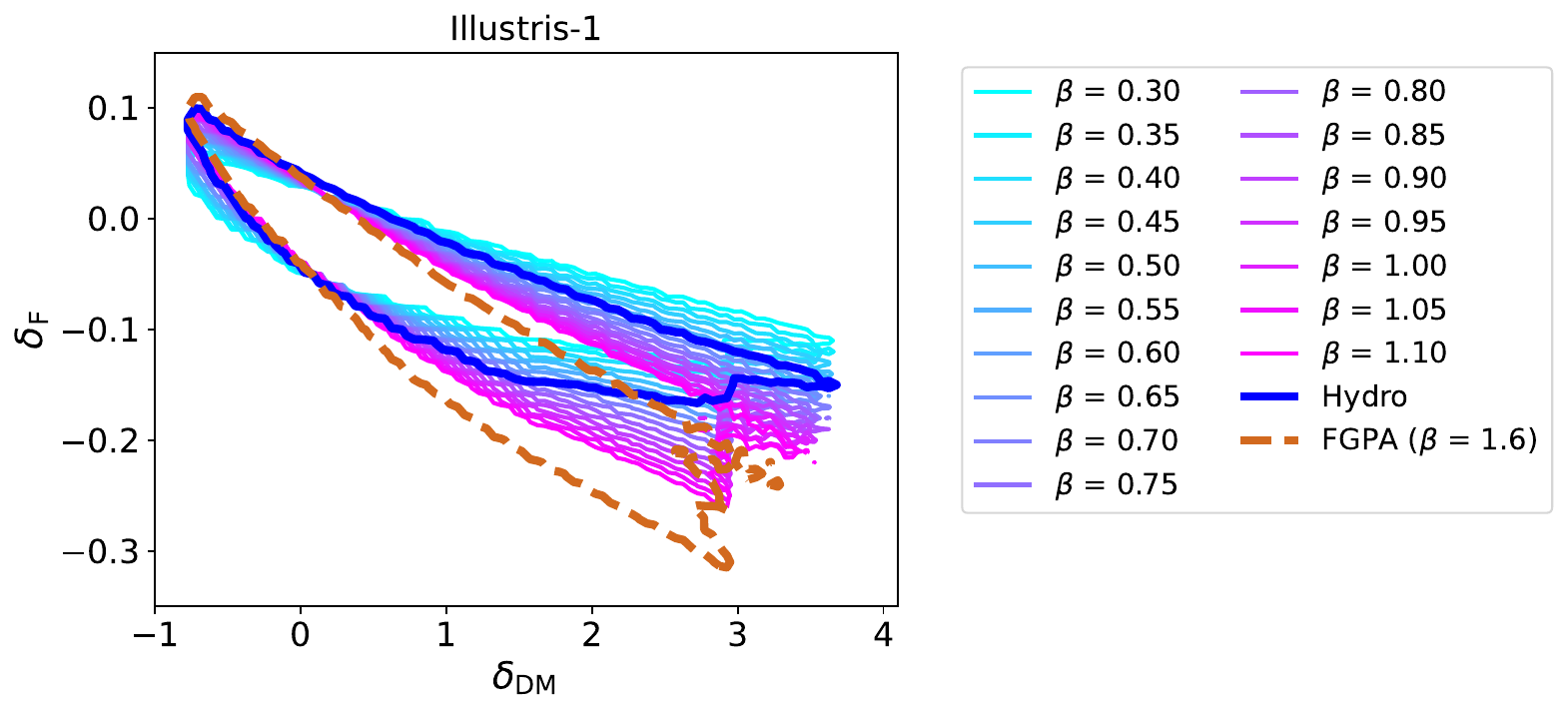}
	\caption{Ly$\alpha$ transmission-DM density distribution for Illustris-1 compared to the FGPA-based distribution as well as the grid with varying FGPA $\beta$ slopes. The contours show the 2\% level of the PDF of the distribution.}
    \label{fig:fitgrid_ill}
\end{figure}

\begin{figure}[hbt!]
    \centering
	\includegraphics[width=\columnwidth]{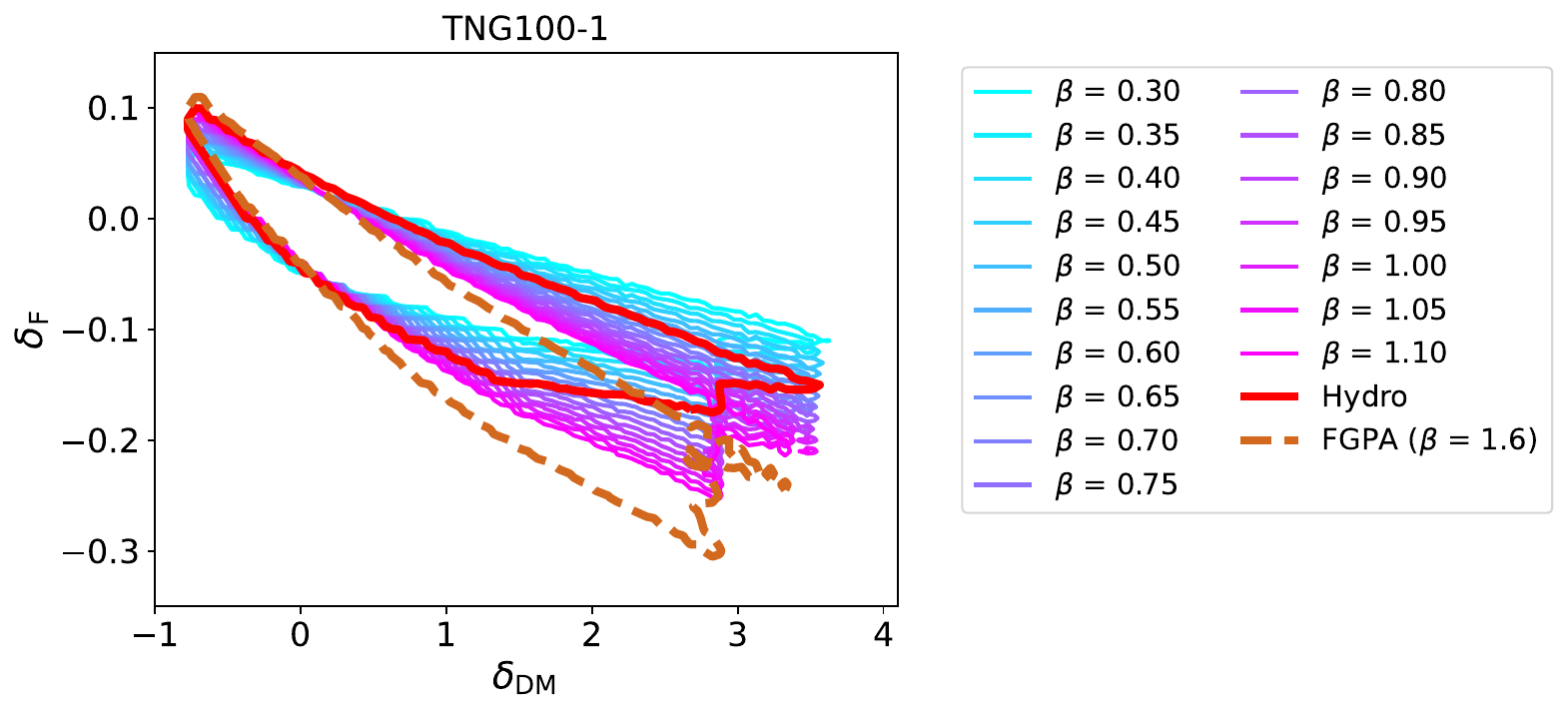}
	\caption{Ly$\alpha$ transmission-DM density distribution for TNG100-1 compared to the FGPA-based distribution as well as the grid with varying FGPA $\beta$ slopes. The contours show the 2\% level of the PDF of the distribution.}
    \label{fig:fitgrid_tng}
\end{figure}

\begin{figure}[hbt!]
    \centering
	\includegraphics[width=\columnwidth]{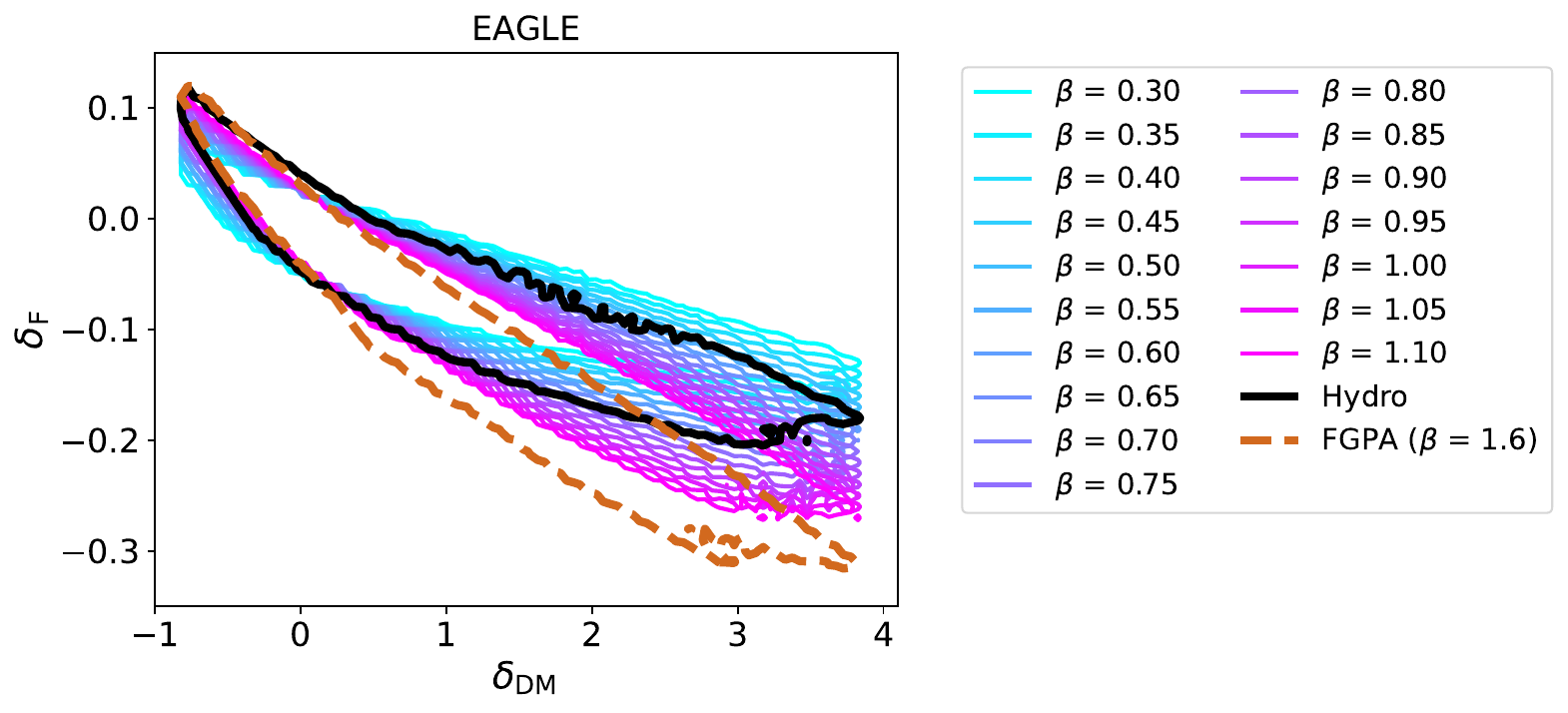}
	\caption{Ly$\alpha$ transmission-DM density distribution for EAGLE compared to the FGPA-based distribution as well as the grid with varying FGPA $\beta$ slopes. The contours show the 2\% level of the PDF of the distribution.}
    \label{fig:fitgrid_eagle}
\end{figure}


\end{document}